\documentclass[a4paper,12pt]{article}
\pdfoutput=1
\usepackage{feynmp-auto,expdlist,multirow}
\usepackage{amsmath, amsfonts, amssymb,eufrak}
\usepackage[mathcal]{euscript}
\usepackage{graphicx}
\usepackage{enumerate}
\usepackage{hyperref}
\usepackage{latexsym}
\usepackage{hepnicenames}
\usepackage{enumerate}
\usepackage{soul}
\usepackage[normalem]{ulem}
 \usepackage{bbold}
\usepackage{units}
\usepackage{pdflscape}
\usepackage[font=small,labelfont=bf,labelsep=period]{caption}
\usepackage{tabularx}
\makeatletter
\def\hlinewd#1{%
\noalign{\ifnum0=`}\fi\hrule \@height #1 %
\futurelet\reserved@a\@xhline}
\makeatother

\usepackage{array,multirow}
\usepackage{boldline}

\newcommand{\A}{{\cal A}}
\newcommand{\B}{{\cal B}}

\newcommand{\C}{{\cal C}}

\newcommand{\ebin}{$(2\text{--}6)$ keVee}

\usepackage{mathrsfs,graphicx,rotating,amsmath,amsfonts,mathtools,booktabs,amssymb,wasysym}
\usepackage{hyperref}\usepackage{slashed}
\usepackage[nosort]{cite}
\usepackage[table,xcdraw,dvipsnames]{xcolor}
\usepackage{bm}
\usepackage{graphicx}
\usepackage{multirow,multicol}
\allowdisplaybreaks

\renewcommand{\[}{\left[}

\newcommand{\mio}[1]{}

\def\bpm{\begin{pmatrix}}
\def\epm{\end{pmatrix}}

\usepackage{mathrsfs}

 \newcommand{\fig}[1]{~\ref{fig:#1}}

\allowdisplaybreaks
\usepackage{multicol}
\usepackage{color}
\definecolor{rosso}{cmyk}{0,1,1,0.4}
\definecolor{rossos}{cmyk}{0,1,1,0.55}
\definecolor{rossoc}{cmyk}{0,1,1,0.2}
\definecolor{blu}{cmyk}{1,1,0,0.3}
\definecolor{blus}{cmyk}{1,1,0,0.6}
\definecolor{bluc}{cmyk}{1,1,0,0.1}
\definecolor{verde}{cmyk}{0.92,0,0.59,0.25}
\definecolor{verdec}{cmyk}{0.92,0,0.59,0.15}
\definecolor{verdes}{cmyk}{0.92,0,0.59,0.4}

\newcommand{\eq}[1]{~{\rm (\ref{eq:#1})}}

\def\circa#1{\,\raise.3ex\hbox{$#1$\kern-.75em\lower1ex\hbox{$\sim$}}\,}

\newcommand{\beq}{\begin{equation}}
\newcommand{\eeq}{\end{equation}}

\newcommand{\bea}{\begin{eqnarray}}
\newcommand{\eea}{\end{eqnarray}}
\newcommand{\be}{\begin{equation}}
\newcommand{\ee}{\end{equation}}
\font\tenrsfs=rsfs10 at 12pt
\font\sevenrsfs=rsfs7
\font\fiversfs=rsfs5
\newfam\rsfsfam
\textfont\rsfsfam=\tenrsfs
\scriptfont\rsfsfam=\sevenrsfs
\scriptscriptfont\rsfsfam=\fiversfs

\newsavebox\MBox

\def\circa#1{\,\raise.3ex\hbox{$#1$\kern-.75em\lower1ex\hbox{$\sim$}}\,}
\makeatletter

\font\ital=cmu10

\def\hhref#1{\href{http://arxiv.org/abs/#1}{arXiv:#1}}
\usepackage{xstring}
\newcommand{\hhrefq}[1]{\IfSubStr{#1}{:}{\href{http://inspirehep.net/search?ln=en&ln=en&p=#1&of=hb&action_search=Search&sf=&so=d&rm=&rg=25&sc=0}{InSpire:#1}}{\hhref{#1}}}

\def\art{\@ifnextchar[{\eart}{\oart}}
\def\eart[#1]#2#3#4#5#6{{\rm #2}, {\em #3 \bf #4} {\rm (#6) #5} ({\em #1})}
\def\article{\@ifnextchar[{\earticle}{\oarticle}}
\def\oarticle#1#2#3#4#5#6{{\rm #1}, {\ital ``#6''}, {\rm #2 #3 (#5) #4}}
\def\earticle[#1]#2#3#4#5#6#7{{\rm #2}, {\ital ``#7''}, {\rm #3 #4 (#6) #5}  [\hhrefq{#1}]}
\def\hepart[#1]#2{{\rm #2, \sl#1}}
\def\heparticle[#1]#2#3{#2, {\ital ``#3''} [\hhrefq{#1}]}
\newcommand{\doi}[1]{\href{http://dx.doi.org/#1}{[link]}}

\newcommand{\hhrefqq}[1]{\IfBeginWith{#1}{10.}{\href{https://doi.org/#1}{doi:#1}}{\hhrefq{#1}}}
\def\earticle[#1]#2#3#4#5#6#7{{\rm #2}, {\ital ``#7''}, {\rm #3 #4 (#6) #5}  [\hhrefqq{#1}]}

\renewenvironment{thebibliography}[1]
     {\begin{multicols}{2}[\section*{\refname}]%
      \@mkboth{\MakeUppercase\refname}{\MakeUppercase\refname}%
      \list{\@biblabel{\@arabic\c@enumiv}}%
           {\settowidth\labelwidth{\@biblabel{#1}}%
            \leftmargin\labelwidth
            \advance\leftmargin\labelsep
            \@openbib@code
            \usecounter{enumiv}%
            \let\p@enumiv\@empty
            \renewcommand\theenumiv{\@arabic\c@enumiv}}%
      \sloppy
      \clubpenalty4000
      \@clubpenalty \clubpenalty
      \widowpenalty4000%
      \sfcode`\.\@m}
     {\def\@noitemerr
       {\@latex@warning{Empty `thebibliography' environment}}%
      \endlist\end{multicols}}

\newcounter{alphaequation}[equation]
\def\thealphaequation{\theequation\hbox to
0.6em{\hfil\alph{alphaequation}\hfil}}
\def\eqnsystem#1{
\def\@eqnnum{{\rm (\thealphaequation)}}
\def\@@eqncr{\let\@tempa\relax \ifcase\@eqcnt \def\@tempa{& & &} \or
  \def\@tempa{& &}\or \def\@tempa{&}\fi\@tempa
  \if@eqnsw\@eqnnum\refstepcounter{alphaequation}\fi
\global\@eqnswtrue\global\@eqcnt=0\cr}
\refstepcounter{equation} \let\@currentlabel\theequation \def\@tempb{#1}
\ifx\@tempb\empty\else\label{#1}\fi
\refstepcounter{alphaequation}
\let\@currentlabel\thealphaequation
\global\@eqnswtrue\global\@eqcnt=0 \tabskip\@centering\let\\=\@eqncr
$$\halign to \displaywidth\bgroup \@eqnsel\hskip\@centering
$\displaystyle\tabskip\z@{##}$&\global\@eqcnt\@ne
\hskip2\arraycolsep\hfil${##}$\hfil& \global\@eqcnt\tw@\hskip2\arraycolsep
$\displaystyle\tabskip\z@{##}$\hfil
\tabskip\@centering&\llap{##}\tabskip\z@\cr}
\def\endeqnsystem{\@@eqncr\egroup$$\global\@ignoretrue} \makeatother

\definecolor{Gray}{gray}{0.95}
\definecolor{darkblue}{rgb}{0,0.1,0.5}
\definecolor{darkgreen}{rgb}{0,0.5,0.2}

\def\bal#1\eal{\begin{align}#1\end{align}}

\def\DAMA{{\sc DAMA}}
\def\LIBRA{{\sc DAMA/LIBRA}}
\def\NaI{{\sc DAMA/}NaI}

\def\ANAIS{{\sc ANAIS}}
\def\COSINE{{\sc COSINE}}

\usepackage{hyperref}
\hypersetup{colorlinks,bookmarksopen,bookmarksnumbered,
linkcolor=darkblue,pdfstartview=FitH,urlcolor=black,citecolor=verde}

\begin{document}
\vspace{1.5cm}

\begin{center}
{\Large\LARGE\Huge \bf \color{rossos}
Annual modulations from secular \\[4mm]
variations:
relaxing DAMA?}\\[1cm]
{\bf Dario Buttazzo$^{a}$, Paolo Panci$^{a,b}$, Nicola Rossi$^{c}$, 
Alessandro Strumia$^{b}$}\\[7mm]

{\it $^a \,$INFN, Sezione di Pisa }
  \\
{\it $^b \,$Dipartimento di Fisica, Universit\`a di Pisa }
\\
  {\it $^c \, $ INFN, Sezione di Roma and Laboratori Nazionali del Gran Sasso } 
  
\vspace{0.5cm}

\begin{quote}
The DAMA collaboration reported an annually modulated rate
with a phase compatible with a Dark Matter induced signal.
We point out that
a slowly varying rate
can bias or even
simulate an annual modulation  
if data are analyzed in terms of residuals computed by
subtracting approximately yearly averages starting  from a fixed date, rather than a background
continuous in time.
In the most extreme case, the amplitude and  phase of the annual modulation reported by \DAMA\
could be alternatively interpreted as a decennial growth of the rate.
This possibility appears mildly disfavoured by a detailed study of the available data,
but cannot be safely excluded. 
In general, a decreasing or increasing rate could partially reduce or enhance a true annual modulation, respectively.
The issue could be clarified by looking at the full
time-dependence of the \DAMA\ total rate, not explicitly published so far.
\end{quote}

\thispagestyle{empty}
\bigskip

\hypersetup{colorlinks,bookmarksopen,bookmarksnumbered,
linkcolor=blus,pdfstartview=FitH,urlcolor=rossos,citecolor=verde}

\end{center}

\setcounter{footnote}{0}
\tableofcontents

\eject


\section{Introduction}\label{Sec:1}

Various experiments are searching for interactions of Dark Matter  (DM) with ordinary matter. 
Most experiments tried to increase the sensitivity to DM collisions
by reducing their backgrounds and found
no evidence for DM. 
The \DAMA\ collaboration followed a different strategy, focused on large statistics: DM scatterings 
would contribute to the total event rate with
an excess which is  annually modulated due to the rotation of the Earth around the Sun.
Indeed, the flux of DM particles hitting the Earth would peak around June, 2$^{\rm nd}$,
when the Earth's orbital velocity is more aligned to the Sun's motion with respect to
the galactic frame.

The \DAMA\ collaboration reported the observation of an annual modulation in single-hit scintillation events in NaI crystals, with the phase expected for a DM signal.
The upgraded \LIBRA\ detector confirmed the earlier result of \NaI,
collecting more data and reaching a significance of about $13\,\sigma$ (loosely speaking)  for the cumulative exposure~\cite{Bernabei:2000qi,DAMA,0804.2741,1308.5109,1805.10486}.  
The \ANAIS~\cite{1812.01377,1903.03973,ANAIS} 
and \COSINE~\cite{COSINE,1905.12861,COSINEvsDAMA} 
experiments employ NaI crystals like  \DAMA, and are also looking for an annual modulation in the counting rate. 
So far they have not reported any modulation, 
but their sensitivity is not yet  sufficient to probe the \DAMA\ signal~\cite{COSINEvsDAMA,1911.12601}. 
Furthermore, a DM signal compatible with the \DAMA\ result 
has not been confirmed by other experiments which use different detectors and techniques (see {\it e.g.}\ the review of DM in~\cite{PDG}). 
These experiments reached high sensitivity to both DM nuclear and electron recoils and, therefore, proposing theoretically viable  DM interpretations of the \DAMA\ detection progressively
becomes very challenging~\cite{hep-ph/0010036,hep-ph/0101138,Fairbairn:2008gz,0912.4264,1107.0715,1108.4661,1110.0857,1203.6652, 1402.1507, 1406.5542, 1602.04074,1703.08942,1804.08437, 1904.07127, 1910.11569}. 

\medskip

The \DAMA\ result remains an open issue. 
Apart from the DM hypothesis, many attempts have been put forward to explain the signal by 
considering speculative oscillating backgrounds with a period of roughly one year (see {\it e.g.}~\cite{0912.2983,1006.5255,1101.5205,1102.0815}). 
Without addressing the plausibility of each proposed background (nor of possible backgrounds that might have been overlooked), the fact that the phase of the observed modulation is close to June, 2$^{\rm nd}$ is considered as a significant argument in favour of the DM interpretation of the signal. Furthermore, the amplitude, time dependence,  and event distribution in the detector array of different oscillating backgrounds cannot satisfy all the peculiar features attributed to the measured annual modulation signal (for details see~\cite{0804.2741} and references therein).

The \DAMA\ collaboration published the cumulative total rate of the single-hit scintillation events as a function of the electron equivalent recoil energy in~\cite{Bernabei:2000qi,0804.2741,1805.10486}.
The full time  dependence was never explicitly presented.
On the other hand, the collaboration presented  the modulated residual as a function of time, computed by subtracting --- roughly every year and roughly starting  from the same date --- the weighted average of the total rate in a given cycle of data-taking. This is a dangerous approach, 
since a time-scale similar to the DM periodicity is introduced in the analysis.
Indeed, the analysis procedure adopted by \DAMA\ transforms
a rate that varies with time into a sawtooth with period given by the duration of the analysis cycles. 
If cycles start around the beginning of September and last roughly one year, a growing background mimics an annual modulation peaked in June. As a result, a slow time-dependence of the total rate, even if not oscillating, becomes a possible source of bias. 
For example, the \DAMA\ modulated amplitude could be generated by a growth of the rate of several percent on a decennial time-scale. 

\bigskip

This paper is organized as follows: in section~\ref{modsec} we describe the general idea, and we show how the amplitude and phase of a modulated signal are related to the time variation of the total rate. 
We apply this to a practical example. 
In section~\ref{DAMA} we briefly present the \DAMA\ analysis. A toy Monte Carlo simulation is performed in~\ref{MC},
showing that a linearly growing rate produces a sawtooth signal that
can appear as a cosine, up to statistical errors.
In section~\ref{DAMAdata} we consider the \DAMA\ data.
We study the detailed time-dependence of the \DAMA\ residuals in section~\ref{fit}, finding that both a cosine and a sawtooth provide acceptable fits to the data, 
with the cosine interpretation being somewhat favoured. 
This extreme possibility is consistent with corollary DAMA studies~\cite{1805.10486},
which include a Fourier analysis.
In section~\ref{fitE} we infer the energy dependence of the secular variation.
We discuss some possible time-dependent backgrounds in section~\ref{bckt}.
Conclusions are given in section~\ref{concl}.


\section{Annual modulation from secular variation}\label{modsec}

In this section we discuss, from a general point of view, how an apparently periodic signal can be mimicked 
by a time-dependent total rate without any modulation. 
Let us consider a total rate $R(t)$ that contains an oscillating signal on top of a 
slowly varying background $R_0(t)$
\beq\label{eq:cos}
R(t) = R_0(t) + \A \cos \left(\frac{2\pi t}{T} - \phi\right).
\eeq
If the rate could be measured with arbitrary precision, 
the modulation would be straightforwardly obtained by directly fitting $R(t)$ to the data.
The phase of its peaks would differ from $\phi$ in the presence of $R_0(t)$.

If instead statistical uncertainties are so large that a single annual period is not clearly visible,
one needs to combine data of multiple periods. Then, a slow time-dependence of  the
background $R_0(t)$ becomes more dangerous.
For constant background, $R_0(t) \equiv \mathcal C$, 
one can consider the average of the rate over any desired time interval $\Delta$, and subtract this quantity from $R(t)$. If $\Delta$ is a multiple of the period $T$ 
the sinusoidal signal averages to zero, and this procedure therefore isolates the signal from the background,
\beq\label{eq:residualC}
\A \cos\left(\frac{2\pi}{T}t - \phi\right) = R(t) - \mathcal C = R(t) - \langle R(t)\rangle_{\Delta}\,.
\eeq
As discussed in section~\ref{DAMA}, the \DAMA\
collaboration followed a procedure along these lines.
However, when $R_0(t)$ is not constant, this procedure
introduces an artificial modulation with period $\Delta$.
As an example, we illustrate our point by considering the simple case where 
the rate varies linearly in time,\footnote{This functional form of the total rate was dubbed `relaxion' in~\cite{1504.07551}.}
\beq\label{eq:Rlinear}
R_0(t)=\mathcal C + \mathcal B t\,. 
\eeq
Subtracting the average of the rate in an interval $\Delta$ centred on $t_0$ yields the residual
\beq\label{eq:sega}
S_0(t) \equiv R_0(t) - \langle R_0(t)\rangle_{\Delta} = \B\left(t - t_0\right),\qquad \text{with\quad $t_0 - \frac{\Delta }{2} < t < t_0 + \frac{\Delta }{2}$ \,,}
\eeq
that becomes a sawtooth wave with period $\Delta$ and amplitude $\B \Delta/2$ if the procedure is repeated over several data-taking intervals of equal length $\Delta$.

The sawtooth can be mistaken for a sinusoid with the same period if its amplitude is small enough as compared with the experimental errors. 
Indeed, fitting a generic periodic function with a sinusoid is equivalent to 
picking up the first term of its Fourier series. 
The Fourier series of eq.~\eqref{eq:sega} in the time interval $\Delta$ is
\beq\label{eq:fouriersaw}
S_0(t) = \mathcal B \Delta \sum_{n=1}^\infty \frac{(-1)^{n+1}}{n\pi} \sin\left(\frac{2\pi n}{\Delta} \, (t-t_0)\right)  \simeq \frac{\mathcal B \Delta}\pi \sin\left(\frac{2\pi}{\Delta} \, (t-t_0)\right) + .... \, ,
\eeq
which can be matched to the form of eq.~\eqref{eq:cos} as
\beq
T = \Delta, \qquad\qquad \A = \frac{\B T}{\pi}, \qquad\qquad \phi = \frac{\pi}{2} + \frac{2\pi t_0}{T} \,.
\eeq
The best-fit sinusoidal wave has an extremum a quarter of period
after the beginning of the sawtooth.
For a decreasing background ($\mathcal B<0$) this extremum at $t_{\rm peak} = T/4$ is a maximum;
for an increasing background ($\mathcal B>0$) it is a minimum, and
the maximum is half a period later, at $t_{\rm peak} = 3T/4$.

The procedure to extract the modulation from the total rate has introduced a bias --- the length $\Delta$ of the time interval over which the background subtraction is periodically applied --- and a nonzero modulation with period $\Delta$ has been generated from a non-periodic rate.
In the presence of a modulated signal as in eq.~\eqref{eq:cos}, its amplitude $\A$ and phase $\phi$ will be modified by the fitting procedure.
In the simple case where the duration of the cycles is taken equal to the period of the signal one has
\beq\label{eq:ABequiv}
\A_{\rm fit}^2 = \A^2 + \frac{\B^2 T^2}{\pi^2} + 2\frac{\A\B T}{\pi}\sin \!\Big(\phi-\frac{2\pi t_0}{T}\Big),\quad 
\tan\phi_{\rm fit} = \frac{\A \sin\phi + ({\B T}/{\pi}) \cos(2\pi t_0/T)}{\A \cos\phi - ({\B T}/{\pi})\sin(2\pi t_0/T)}.
\eeq

\subsection{Extracting a modulation from data: a bibliometric example}

\begin{figure}[t]
\includegraphics[height=5.2cm]{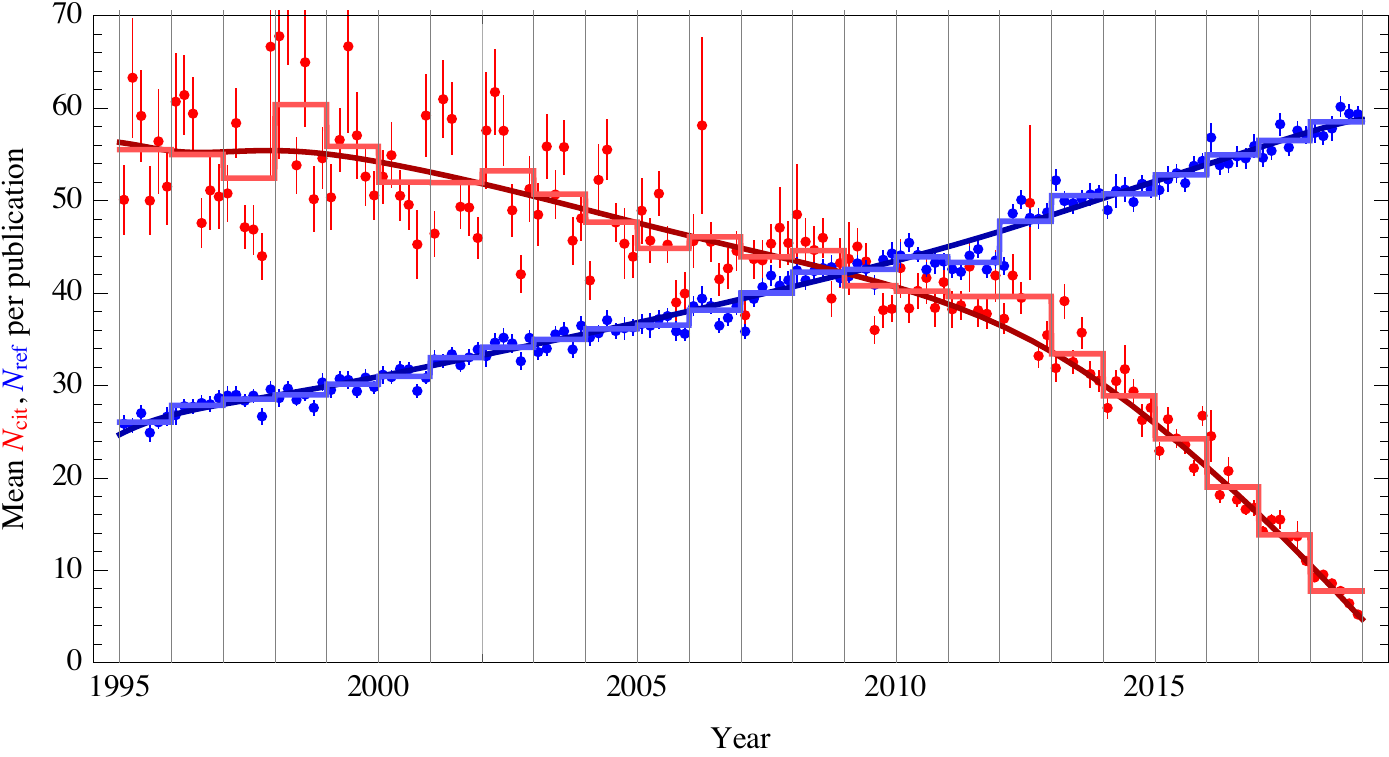}\hfill%
\includegraphics[height=5.2cm]{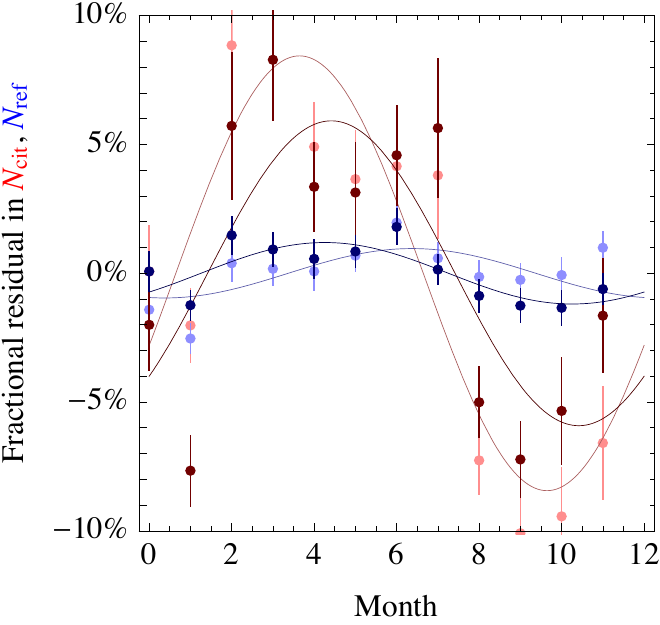}
\caption{\label{fig:biblio}
\em {\bf Left:} Average number of {citations $N_{\rm cit}$ (red)} and of 
{references $N_{\rm ref}$ (blue)}  of publications in high-energy physics
in arXiv as function of time.  
Both show an annual modulation together with a trend on a slower, decennial, time-scale:
$N_{\rm cit}$ is decreasing and $N_{\rm ref}$ is growing (smooth curves).
{\bf Right}: Monthly variation computed integrating over the years 
and subtracting the slow trend in two different ways: 
subtracting the  discontinuous yearly averages (lighter red and blue)
or their smooth averages (darker red and blue).}
\end{figure}

In order to work out a practical example we consider real data from bibliometrics.
We consider all publications in high-energy physics (excluding astro-ph) since 1995, and we compute the bi-monthly averages
of their number of references $N_{\rm ref}(t)$, and of the number of citations $N_{\rm cit}(t)$
they received up to now, as reported in the {\sc InSpire} database~\cite{InSpire}. The data are shown in the left panel of fig.~\ref{fig:biblio} as a function of publication time $t$.
The average number of references per paper is slowly growing with time because of the overall expansion of the field.
The number of citations is slowly decreasing with time because more recent papers have not yet fully accumulated citations.
These slow variations produce an apparent annual modulation with fractional amplitude
\beq \frac{\cal A}{N} \approx \frac{\dot N}{N} \frac{1\,{\rm yr}}{\pi} \approx 1\%\,, \eeq
when data are analyzed following the procedure described above ({\it i.e.} by subtracting the yearly averages from $N(t)$).
The artificial modulation peaks approximately 3 months later than the start of the analysis period (January, 1$^{\rm st}$) for citations, since $N_{\rm cit}(t)$ grows with time, and approximately 9 month later for references, since $N_{\rm ref}(t)$ decreases with time.

A different analysis procedure allows to search
for a modulation over a slowly varying background
without introducing an artificial modulation that biases the signal.
The residuals are now computed by subtracting a smooth function that follows the average slow
evolution of the background, rather than the discontinuous yearly averages.
Such functions can be obtained, for example, as a 1$^{\rm st}$ or 2$^{\rm nd}$ order interpolation of the yearly averages.
In this language the `dangerous' procedure corresponds 
to subtracting a 0$^{\rm th}$ order discontinuous interpolation.
The two possibilities are plotted in lighter/darker color in fig.\fig{biblio}.
The right panel of  fig.~\ref{fig:biblio} shows the
residuals computed following the two procedures.
As expected, the two modulation residuals differ by a few $\%$, and their phase is shifted.
The  correct procedure shows that $N_{\rm cit}$ oscillates by about $5\%$ and
that $N_{\rm ref}$ oscillates by about $1\%$; both peak around April.\footnote{This is, by itself, an original new contribution to bibliometrics. A modulation is present because
duties, conferences, jobs applications follow a periodic calendar.
The modulation amplitude becomes much larger and visible by eye
if unpublished papers and conference proceedings are included.}
The `dangerous' procedure incorrectly determines the modulation amplitudes and their phases.


\section{The DAMA analysis: simulation}\label{DAMA}

We now apply the above considerations to the \DAMA\ analysis. The  experiment is looking for  an annual modulation in the rate of  single-hit scintillation events  in NaI crystals, as a possible signal of DM interactions with matter. The total  rate is dominated by large backgrounds due to natural radioactivity of the detector and surrounding environment. As discussed above, a background that evolves slowly on a longer time-scale (say decennial) can simulate an annual modulation if the analysis is performed by subtracting yearly weighted averages.
Since this is the procedure followed by the  \DAMA\  collaboration, we study the possible impact of this observation on the results. 

The cumulative total rates 
expressed in cpd/kg/keVee 
(where `ee' stands for electron equivalent)
have been presented as a function of the electron-equivalent recoil energy in three long-run phases in fig.~1 of~\cite{Bernabei:2000qi}, \cite{0804.2741}, and \cite{1805.10486} for \NaI, \LIBRA\ Phase~1, and \LIBRA\ Phase~2, respectively. 
A change of the total rate between the three different phases is evident, and is due to improvements in the experimental setup causing an overall background reduction. However, its time dependence within a single data-taking phase is not provided by the  collaboration, and cannot be inferred from these data alone.

The time dependence is reported for the residuals. In each of the three long-run phases, the \DAMA\ residuals are derived from the measured rate of the single-hit scintillation events, after subtracting the time average of the rate over each cycle (of roughly yearly duration) such that ``the weighted mean of the residuals must obviously be zero over one cycle'' (see~\cite{Bernabei:2000qi, 0804.2741, 1308.5109} and references therein). 
As discussed in the previous section, if the rate has the form of eq.\eq{cos} with a time-independent background $R_0(t) \equiv \mathcal C$, this procedure correctly subtracts the constant term. 
If instead $R_0(t)$ is slowly varying, the secular term contributes to the residuals, getting transformed to a sawtooth with periodicity equal to the duration of the \DAMA\ cycles (injected because the analysis procedure is repeated quite regularly every year).
The \DAMA\ cycles start every year roughly around the beginning of September, so a growing background can appear like a modulated amplitude peaked at the beginning of June. On the contrary, the peak of the modulation would be around the beginning of December for a decreasing background.

\medskip

We now show that a linearly growing background $R(t) $ without a periodic term can mimic a modulation that can look like the \DAMA\ signal. 
We do so by first performing a Monte Carlo simulation, and next
computing the residuals of the artificial data following the same procedure adopted by \DAMA.

\subsection{Monte Carlo simulation}\label{MC}

The Monte Carlo simulation is done by employing a setup similar to the \LIBRA\ detector. In particular, for both Phase~1 and Phase~2 the total detector mass is taken to be $M = 242.5$ kg (25 NaI crystals with a mass of 9.7 kg each) and we generate events in the energy window $(2\text{--}\,6)$  keVee such that $\Delta E = 4$ keVee. The exposure of the 7 data-taking cycles in Phase~1 is 1.03~$\rm ton \cdot yr$, and their total duration is $\Delta T_1 = 2560$ days (table~1 of~\cite{1308.5109}); the exposure and total duration of the 6 data-taking cycles in Phase~2 are 1.13~$\rm ton \cdot yr$ and $\Delta T_2 = 2153$ days (table~1 of~\cite{1805.10486}). As a consequence, we assume a constant duty cycle efficiency $\zeta = {\rm exposure}/(M \Delta T$) of $60\%$ and $80\%$ for Phase~1 and Phase~2, respectively. 

\begin{figure}[t!]
\hspace{6.pt} \includegraphics[width=0.981\textwidth]{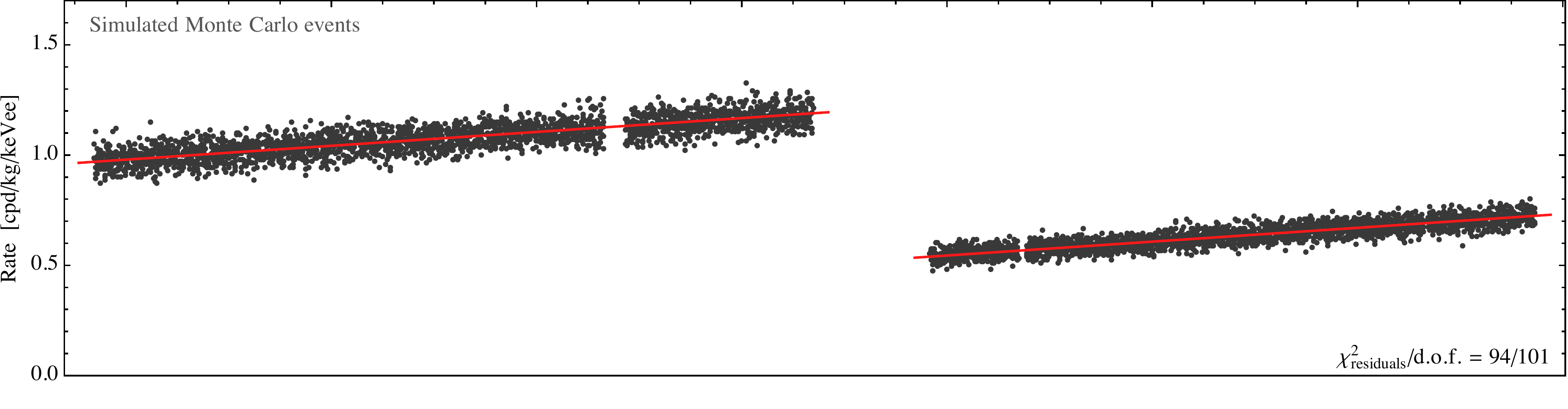}\vspace{-11.1pt}
\includegraphics[width=\textwidth]{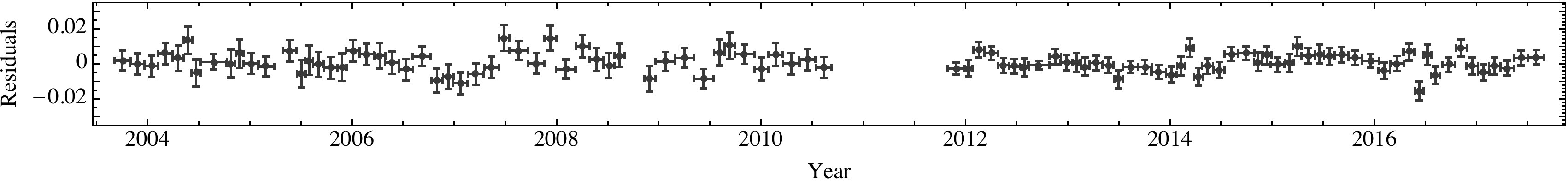}
\caption{\em {\bf Upper}:
Monte Carlo events generated assuming 
no modulation, a slow linear growth in time, $\B = (0.01\,\pi/{\rm yr})\,{\rm cpd/kg/keVee}$, and the same cumulative rate as \LIBRA\ Phase 1 and 2.
{\bf Lower}: Residuals computed with respect to the smooth average
are compatible with no modulation.
\label{fig:fakedata2}}
\end{figure}

We generate a number of random events for each day following a Poisson distribution with mean
\beq\label{eq:Nfake}
N(t) = \zeta M  \Delta E \, R_0(t)  \qquad \mbox{with } \qquad R_0(t) =  \C + \B t .
\eeq
We assume that $R_0(t)$ grows linearly with time with a coefficient $\B = \A \pi /\rm yr$, where $\A = 0.01 \, \rm cpd/kg/keVee$ is roughly  the amplitude of the \DAMA\ modulated signal.
The constant $\C$ is chosen such that the time average of $R_0$ over the whole data-taking period
is equal to the \DAMA\ cumulative total rate in the \ebin\ energy window: for Phase 1, $\langle R_0\rangle_1 = 1.03$ cpd/kg/keVee in an exposure of 0.53 $\rm ton \cdot yr$~\cite{0804.2741}\footnote{In this case  the cumulative total rate is only reported in an exposure of 0.53 $\rm ton \cdot yr$ corresponding to 4 cycles of data-taking that last 1406 days~(table 1 of~\cite{0804.2741}). As a consequence the average of $R(t)$ is performed over a time interval shorter than the total duration of Phase 1, $\Delta T = 2560$ days.}; for Phase 2, $\langle R_0\rangle_2 = 0.63$ cpd/kg/keVee in an exposure of 1.13 $\rm ton\cdot yr$~\cite{1805.10486}. 
The upper panel of fig.~\ref{fig:fakedata2} shows our simulated rate in cpd/kg/keVee as a function of time. The red lines represent the assumed  backgrounds $R_0(t)$ in Phase~1 (on the left) and Phase~2 (on the right). 
Some short periods without data points are included consistently with the \LIBRA\ data acquisition.

\begin{figure}[t]
\includegraphics[width=\textwidth]{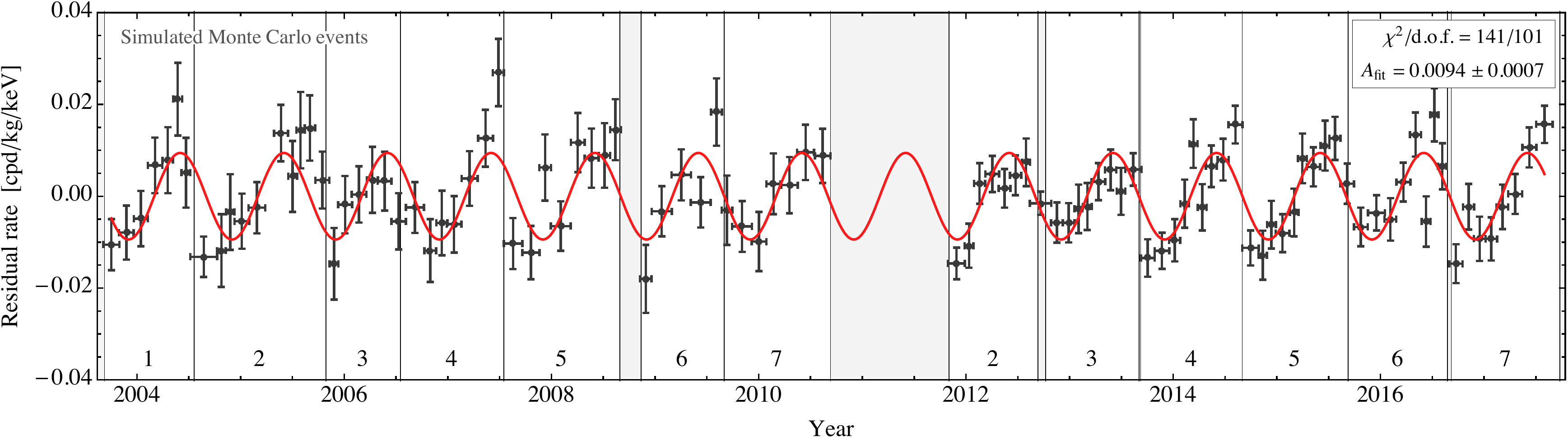}
\caption{\em Residuals
obtained from the simulated data of fig.\fig{fakedata2} (black points)
and computed within each data-taking cycle (vertical lines)
following the \DAMA\ procedure.
The red curve shows the best fit to the simulated residuals with
a DM cosine signal with a period of one year and peaked on June, 2nd. 
The zero-signal hypothesis is excluded with a significance of $12.7\,\sigma$
despite that no modulation was assumed in the simulated data.
\label{fig:fakedata}}\end{figure}

We now compute the residuals of the simulated events following the procedure adopted by \DAMA, {\it i.e.}\ subtracting the average event rate in each cycle. As expected we get a sawtooth up to statistical errors. A complication with respect to the ideal analysis of section~\ref{modsec} arises 
because the \DAMA\ data-taking cycles have slightly irregular durations. The start and end dates of the 7 cycles of Phase~1 and the 6 cycles of Phase~2 can be found in table~1 of~\cite{1308.5109} and~\cite{1805.10486} respectively. Subtracting the average rate in the $k$-th cycle that lasts $\Delta T_k = t_{f,k} - t_{i,k}$ the residuals of eq.\eq{Nfake} follow the irregular sawtooth 
\beq\label{eq:Sfake}
\mathcal S(t) = \zeta M  \Delta E \, S_0(t)  \qquad \mbox{with } \qquad
S_0(t) = 
\mathcal B \left(t - \frac{t_{i,k} + t_{f,k}}{2}\right) \quad \text{ for $t_{i,k} < t < t_{f,k}$} \, ,
\eeq
which is no longer a perfectly periodic function. We collect the residuals in 102 time bins of approximately 1.5 months each, adopting the same binning as the one used in the \DAMA\ analysis. Fig.\fig{fakedata} illustrates the binned residuals expressed in cpd/kg/keVee. The errors are 
of the same order of the errors of the \DAMA\ residuals in the $(2\text{--}\,6)$ keVee energy window.

This procedure results in
something that looks like a cosine annual modulation.
Unlike a true sinusoidal modulation, discontinuities are present between the various cycles. However, the binning procedure can partially wash out these discontinuities:
if a time bin falls in two different cycles, as sometimes happens for the \DAMA\ binning, the sawtooth will approximately average to zero over the bin, and the residual number of events in the bin will be small.
Averaging the rate over 25 crystals with different efficiencies and duty cycles could also have an impact in this respect.

The consistency of the residual rate with a DM signal can be quantified by fitting the residual rate with a cosine with a period $T = 1$ year, and peaked at $t_{\rm peak} = 152.5$ days which corresponds to June, 2$^{\rm nd}$. For our simulated Monte Carlo sample, the fit results in a cosine amplitude $\A = (0.0094\pm 0.0007)\,{\rm cpd/kg/keVee}$, consistent with the injected value of 0.01 cpd/kg/keVee. The goodness of the fit is given by $\chi^2 = 141$ for 101 degrees of freedom. 
The cosine signal is preferred over the zero-signal hypothesis at $12.7\,\sigma$, 
despite that no cosine modulation was present in the simulated data.

If instead the residuals are recomputed by subtracting
the average rate as fitted to a linear function of time over the full data-taking period,
the residuals  (shown in the lower panel of fig.~\ref{fig:fakedata2})
are consistent with no modulation within one sigma ($\chi^2/{\rm d.o.f.} = 94/101$).

\begin{figure}[t]
\includegraphics[height=5.3cm]{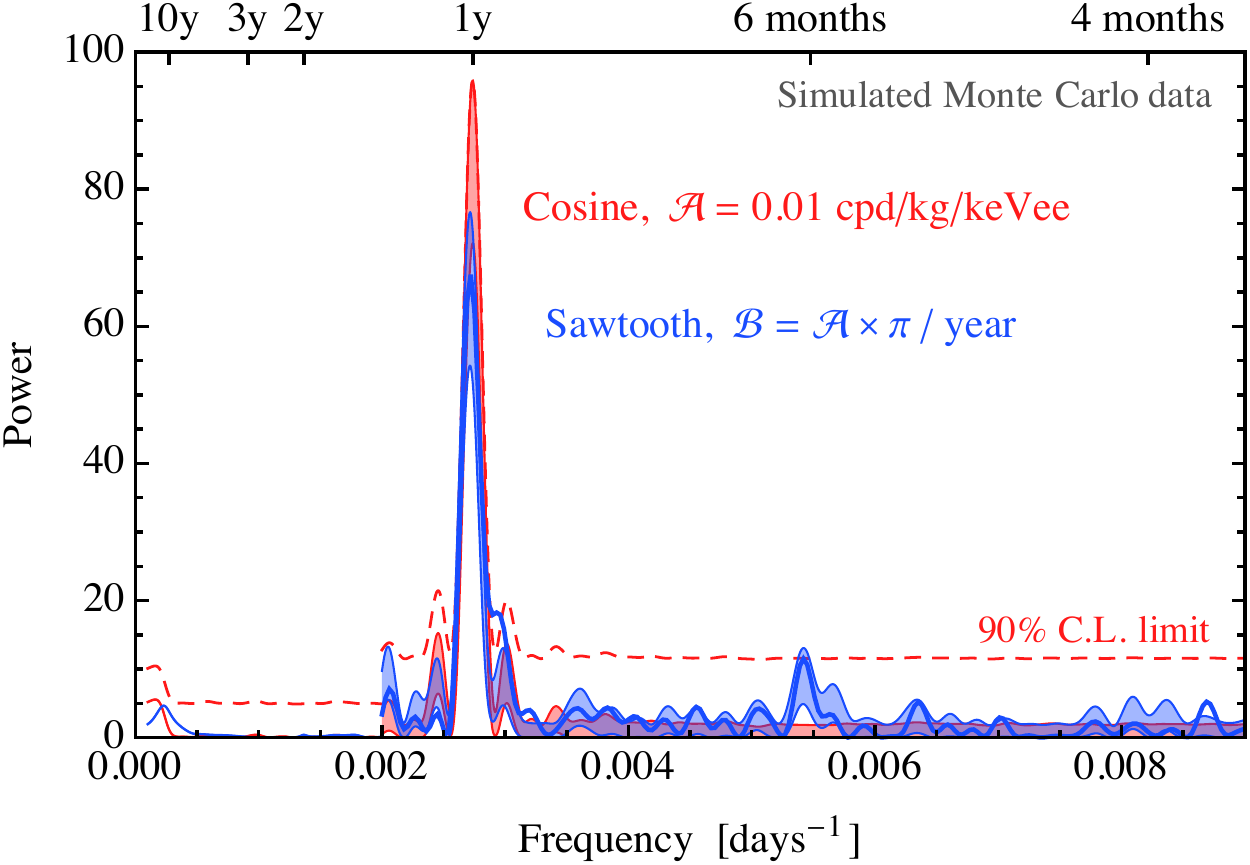}\hfill
\includegraphics[height=5.3cm]{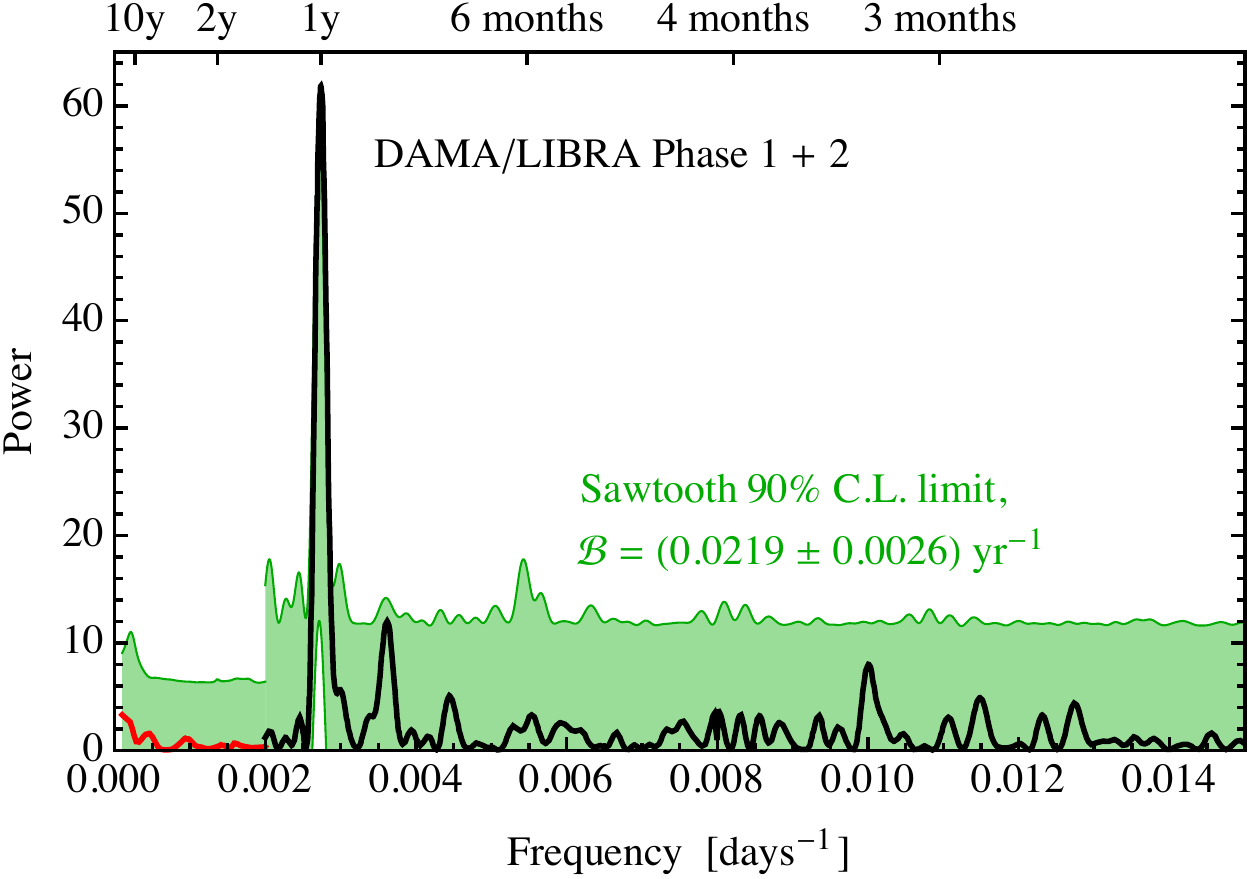}
\caption{\label{fig:power}
\em {\bf Left:} Power spectra computed from Monte Carlo simulations assuming 
a sawtooth (blue line and band) and a cosine modulation (red band), with $\A = \B\cdot {\rm yr}/\pi = 0.01\, {\rm cpd/kg/keVee}$.
The colored bands represent local $1\sigma$ intervals. The red dashed line shows the global 90\% C.L.\ limit for a sinusoidal signal.
{\bf Right:} Power spectrum of the \DAMA\ residuals (black) and annual rate (red) for Phase~1 and Phase~2 combined, taken from \cite{1805.10486}, together with the 90\% C.L. allowed region (green) obtained from Monte Carlo simulations assuming a sawtooth with $\B = 0.0219\pm 0.0026\,\,{\rm cpd/kg/keVee/yr}$ as in the combined \LIBRA\ best fit of Table~\ref{tab:fittoDAMA}.
}
\end{figure}

In the left panel of fig.\fig{power} we show the power spectrum
(Lomb-Scargle periodogram as defined in~\cite{hep-ph/0605212}) 
corresponding to data
simulated assuming a sawtooth as in fig.\fig{fakedata2} (blue), and compare it to the power spectrum of a
simulated annual cosine modulation (red) with $\A = \B\cdot{\rm yr}/\pi = 0.01$ cpd/kg/keVee.
We computed the power spectrum 
following the procedure adopted by the DAMA collaboration in~\cite{1805.10486}.
In the figure, the colored regions correspond to local $1\sigma$ intervals, computed by simulating 100 Monte Carlo samples; the red dashed line shows the global 90\% C.L.\ allowed region, where all the peaks are expected to fall under the assumption of a sinusoidal signal of given amplitude (eq.~(4) of~\cite{hep-ph/0605212}). We observe that:
\begin{itemize}
\item[$\diamond$] The power spectra at frequencies around and above 1/yr
are obtained from the {\em residuals}
computed within each cycle, and collected in bins of one day.
The number of data points used to obtain the spectra is 4341 days, to match the value used in~\cite{1805.10486}. 
We see that both the sawtooth and the cosine produce comparable peaks at $\nu \approx 1/{\rm yr}$.
The sawtooth contains extra Fourier modes with frequency
$\nu \approx n/{\rm yr}$ and $n$ integer:
the corresponding power is small
because their amplitudes scale as $1/n$ (see eq.~\eqref{eq:fouriersaw}), and the
cycles are not exactly annual.
\item[$\diamond$] Following~\cite{1805.10486}, the power spectra at low frequencies below 1/yr 
are computed from average {\em event rates} in each (roughly annual) cycle.
The main low-frequency peaks are due to the change in the total rate between Phase~1 and Phase~2; they do not have a statistically significant power because they are based on 7+7 time bins only.
\item[$\diamond$] A power spectrum computed from daily total event rates was not
considered in~\cite{0804.2741,1308.5109,1805.10486}.
In the sawtooth simulation, it would exhibit significant power at low frequencies.
\end{itemize}
To summarize, a sawtooth modulation resulting from taking yearly residuals of a linearly growing background is statistically compatible with a sinusoidal modulation, given the available data. This conclusion is obtained both by fitting the explicit time series, and by performing a frequency analysis.


\section{The DAMA analysis: experimental data}\label{DAMAdata}

We now consider the real data collected by the \DAMA\ experiment.
The total rate as a function of time has not been explicitly published, and therefore we cannot directly know whether a secular evolution might have affected the determination of the modulation amplitude or its phase. 
We hence consider the published \DAMA\ residuals, 
and show that their detailed time dependence is consistent
even with the most extreme possibility:
a sawtooth originating from a slowly growing background.
A superposition of cosine and sawtooth, with the main contribution coming from the sinusoidal signal, is however preferred by a global fit.

\begin{figure}[t]
\includegraphics[width=\textwidth]{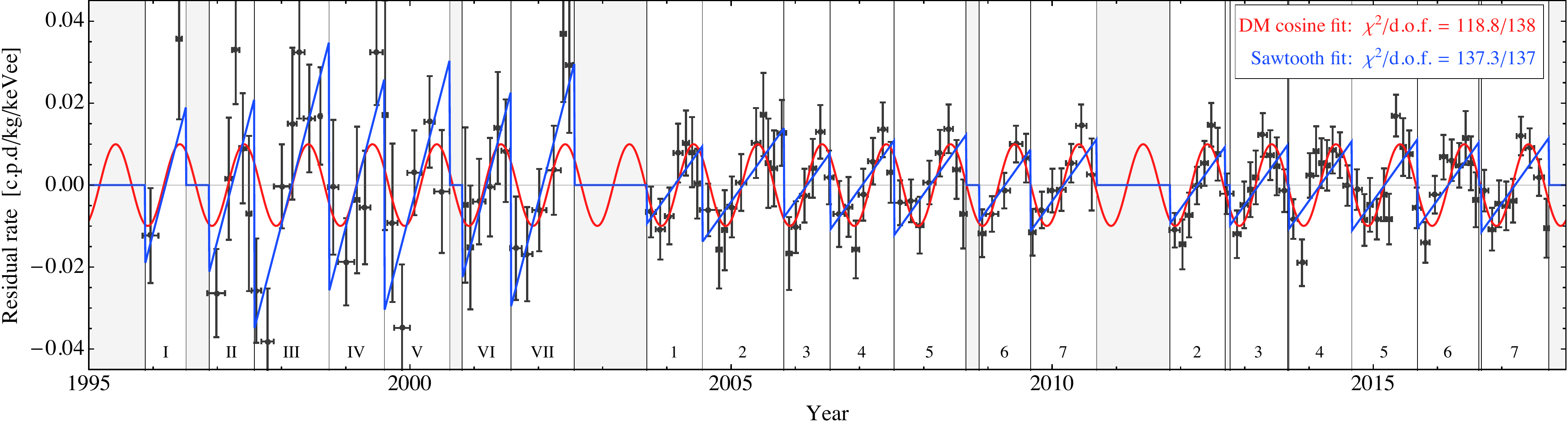}
\caption{\label{fig:DAMAdata_26} 
\em The black data points are the \DAMA\ residuals in the $(2\text{--}\,6)$~${\rm keVee}$ energy window, taken from \cite{Bernabei:2000qi,1805.10486}.
The curves are fits to a cosine annual modulation peaked on June, 2nd (red curve), as expected for a DM signal,
and to the irregular sawtooth obtained from a continuously growing background (blue curve). 
The roughly annual
data-taking cycles of {\rm \NaI}, \LIBRA\ Phase~1, and \LIBRA\ Phase~2 are shown as vertical lines.}
\end{figure}

\subsection{Fit to DAMA residuals}\label{fit}

\begin{figure}[t]
\includegraphics[height=7.3cm]{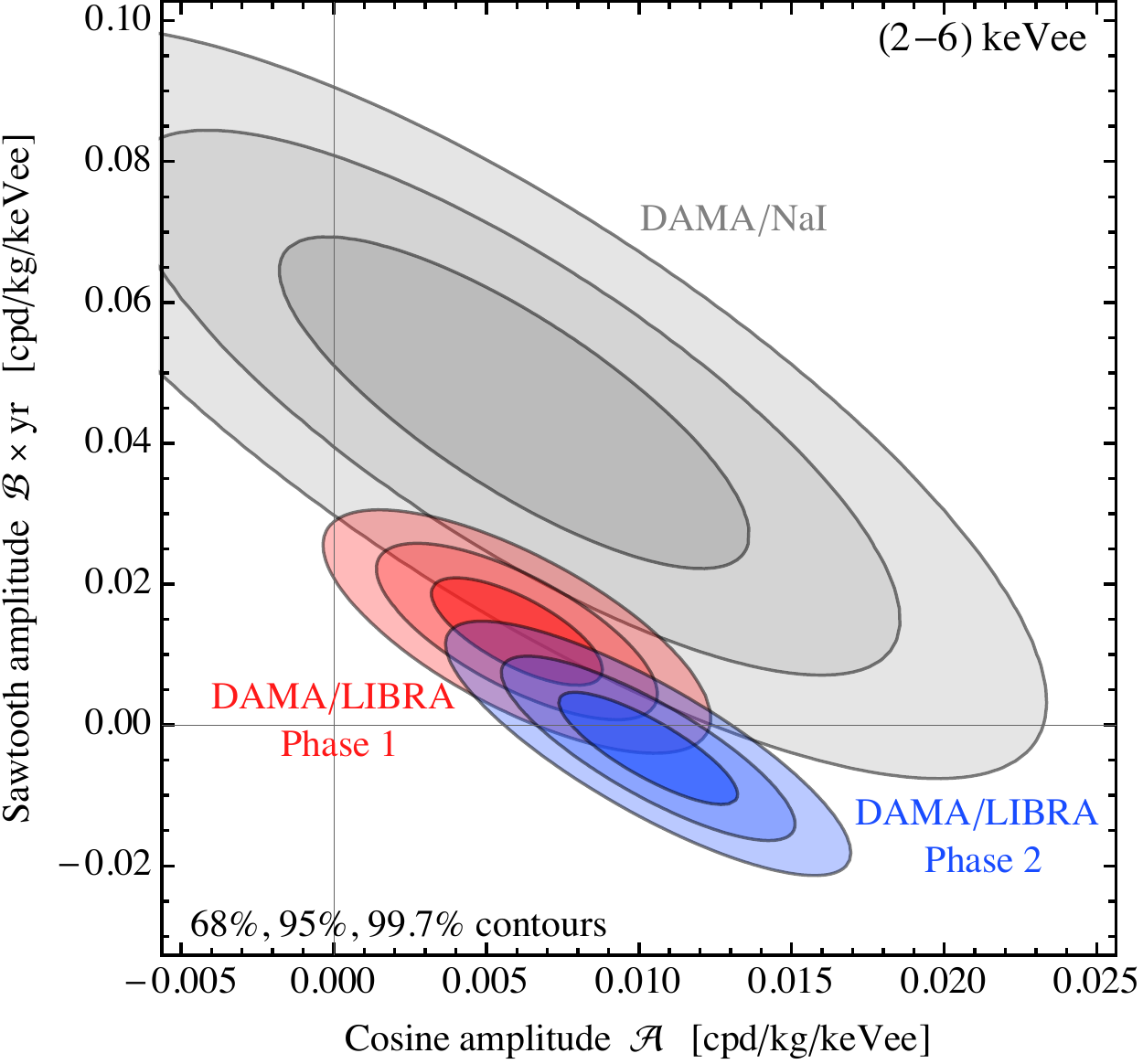}\hfill%
\includegraphics[height=7.3cm]{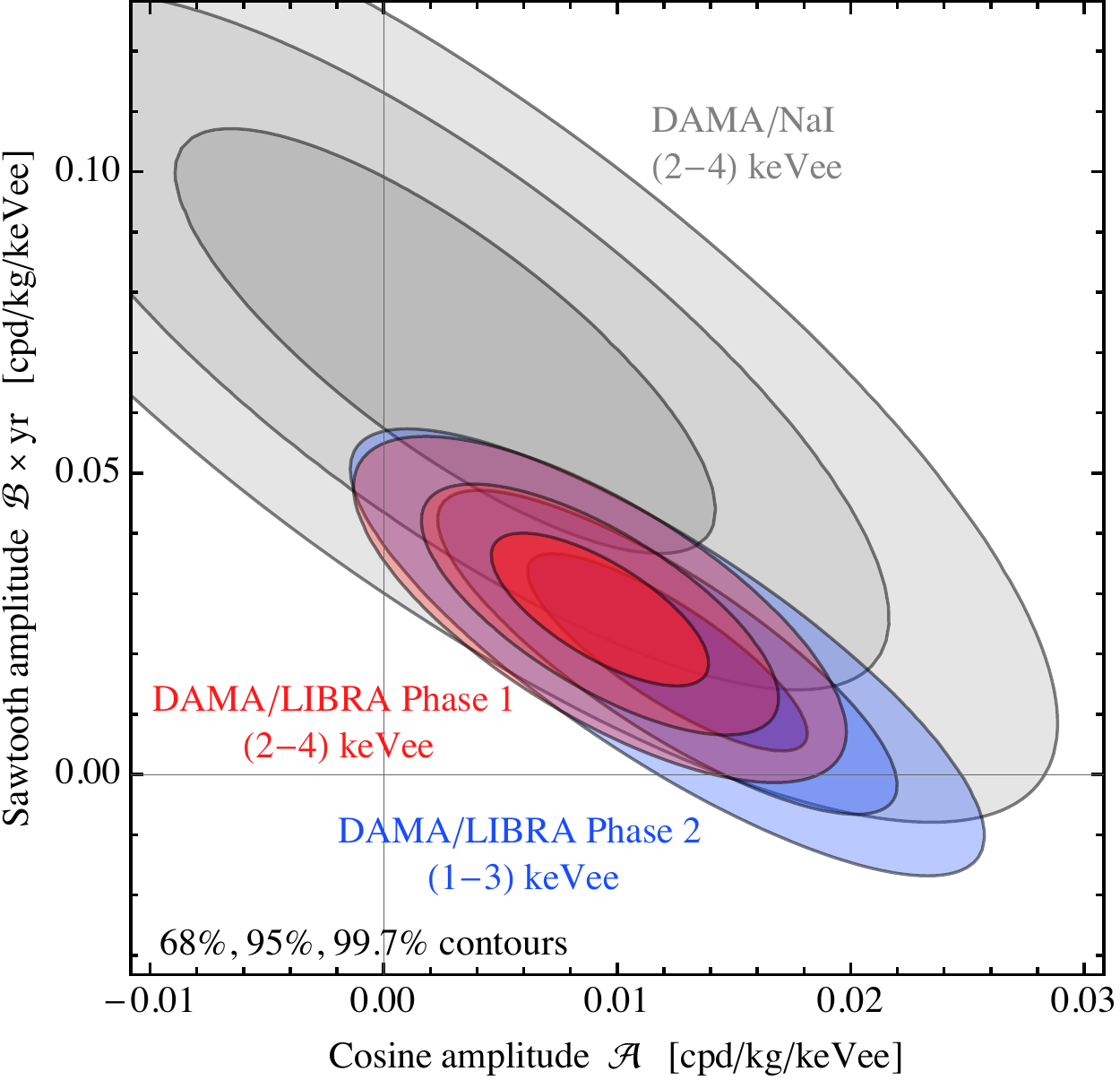}
\caption{\label{fig:DAMAfit_AvsB}
\em Fit of the DAMA residual rates to a sawtooth plus a cosine with period and phase as predicted by DM.
{\bf Left:} fit of residuals in the $(2\text{--}\,6)$~{\rm keVee} energy window for {\rm \NaI} (gray), \LIBRA\ Phase 1 (red) and Phase 2 (blue).
{\bf Right}: fit of residuals in the $(2\text{--}\,4)$ {\rm keVee} energy window for {\rm \NaI} (gray) and \LIBRA\ Phase 1 (red), and in the $(1\text{--}\,3)$ {\rm keVee} energy window for \LIBRA\ Phase 2.}\end{figure}

We again assume a linearly growing total rate as in eq.~(\ref{eq:Rlinear}). This assumption is motivated by the amplitude of the observed modulation seeming constant over the 
Phase 1 and Phase 2 cycles, which requires a slow constant growth of the rate if the modulation is interpreted as a sawtooth.
Furthermore, a constant growth is a good approximation if the time-scale of the secular evolution is  long compared to the duration of the data-taking cycles. However, the time dependence of the background is unknown, and details of our analysis depend of course on it.

The experimental residual rates as a function of time in three different energy intervals are taken from~\cite{0804.2741}, \cite{1308.5109} and \cite{1805.10486} for \NaI, \LIBRA\ Phase~1 and Phase~2 respectively. The residual rate in the $(2\text{--}6)$ keVee energy interval is extracted from~\cite{0804.2741, 1805.10486} and  shown in fig.\fig{DAMAdata_26} as black points, together with the timespan of the 20 annual cycles of the experiment. We fit all these data points with the following two signal functions:
\begin{enumerate}
\item A cosine modulation $\A \cos [2\pi (t -t_{\rm peak})/T]$ with
$T = 1\,{\rm yr}$ and $t_{\rm peak} =$ June, 2$^{\rm nd}$, 
as predicted for a DM-induced annual modulation signal;
\item The irregular sawtooth generated by a growing background $\mathcal{B} t$, taking into account the details of the \DAMA\ cycles as in eq.~\eqref{eq:Sfake}.
\end{enumerate}

In the fit we allow for two different values of the sawtooth amplitude:  $\mathcal{B}_{\rm LIBRA}$ for the two phases of \LIBRA, and $\mathcal{B}_{\rm NaI}$ for \NaI, as both the experimental apparatus and the total mass of the crystals were different.  The best-fit curves for the two cases are shown in fig.~\ref{fig:DAMAdata_26}. We find that both the cosine and the sawtooth provide good fits to the data,
as they have a $\chi^2$ per degree of freedom of order one. 
More precisely, the sawtooth fit has $\chi^2_{\rm saw}/{\rm d.o.f.} = 137.1/137$,
and the cosine fit has $\chi^2_{\cos}/{\rm d.o.f.} = 118.8/138$.\footnote{We follow the counting
of degrees of freedom adopted in \DAMA\ papers, although the time bins within any given cycle 
can get correlated by the analysis procedure.}

The best-fit value for the cosine amplitude is in perfect agreement with the result reported by the DAMA collaboration, $\A = (0.0102\pm 0.0008)$ cpd/kg/keVee~\cite{1805.10486}.
 We get similar results when leaving the phase and period of the cosine free to vary.
For the sawtooth, the best-fit values of the $\B$ coefficients are
\begin{eqnsystem}{sys:Bfit}
\B_{\rm NaI} &= (0.060\pm 0.009)\, {\rm cpd/kg/keVee/yr}\,,\\
\B_{\rm LIBRA} &= (0.022\pm 0.003)\, {\rm cpd/kg/keVee/yr}\,,
\end{eqnsystem}
for \NaI\ and \LIBRA, respectively, which correspond to a yearly growth of the total rate of a few percent. Allowing for different coefficients for Phase~1 and Phase~2, and even for each individual cycle, gives similar results. 

\medskip

The DAMA collaboration also reports residual rates in different energy windows.
We perform a fit in all the energy intervals of \NaI, \LIBRA\ Phase~1 and Phase~2, and we find that both the cosine and the sawtooth give acceptable fits to the data.  
The two theoretical hypotheses result in similar $\chi^2/{\rm d.o.f.}$ in all cases, except for the higher energy bin of \LIBRA\ Phase~2 where a cosine is favoured. We also provide combined results with a single $\B$ for the three phases in the $(2\text{--}\,6)$  keVee energy interval where this is possible. The results of all our fits are summarized in table~\ref{tab:fittoDAMA}.

\begin{figure}[t]
\includegraphics[height=7.5cm]{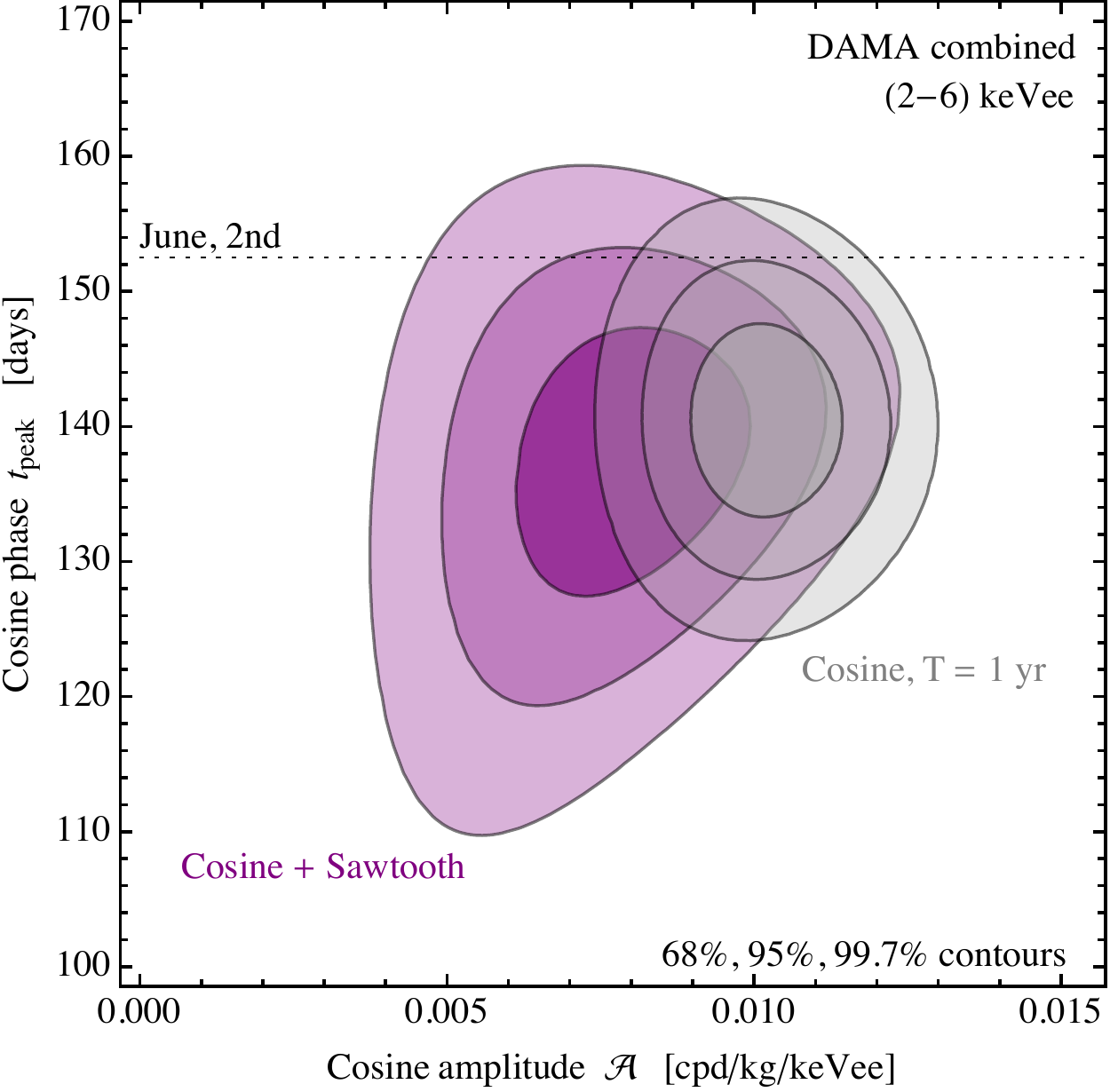}\hfill%
\includegraphics[height=7.5cm]{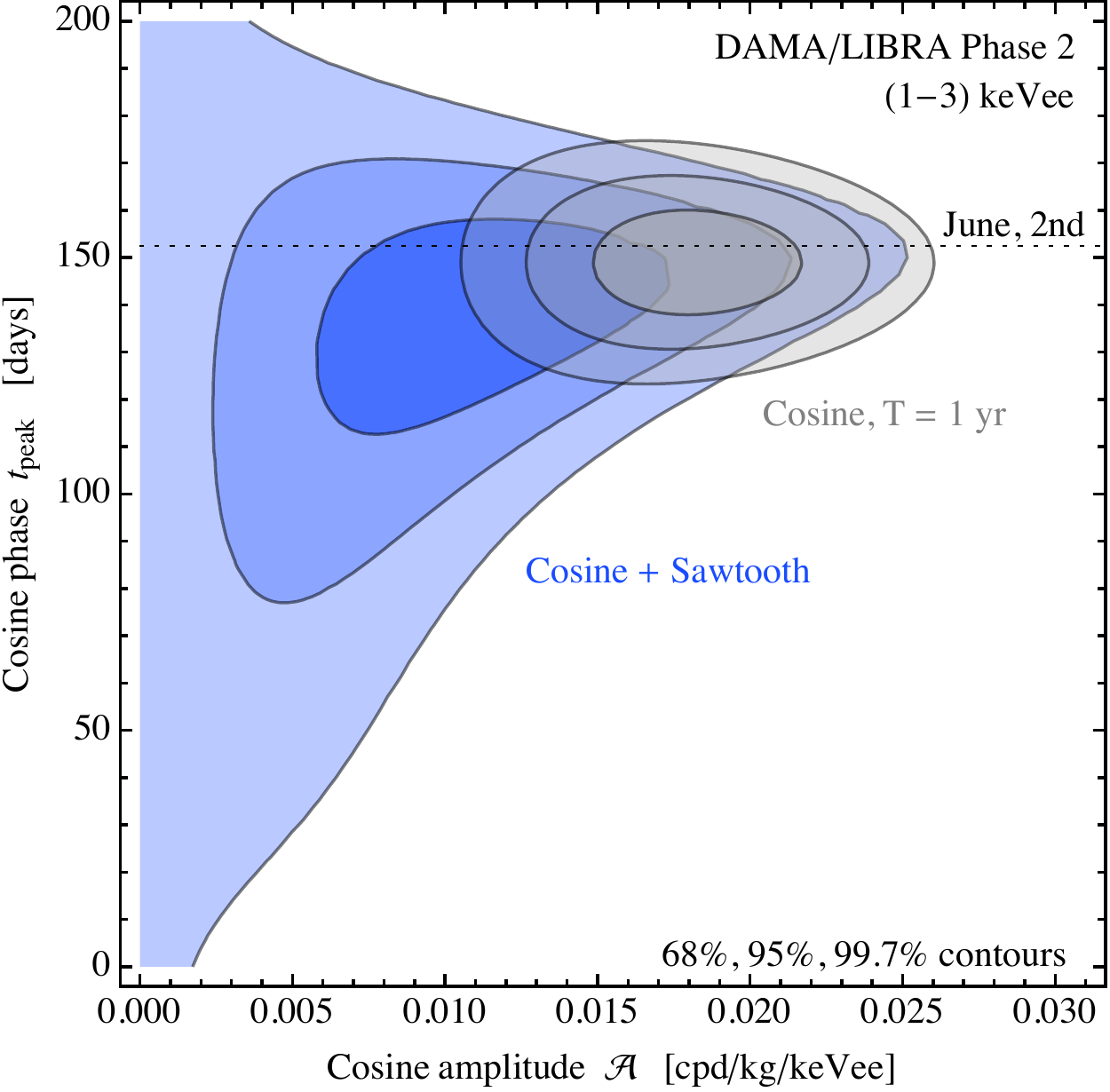}
\caption{\label{fig:DAMAfit_AvsPhi}
\em {\bf Left:} Fit of the cosine amplitude and its phase to \DAMA\ residuals, allowing for a sawtooth component (colored regions) or setting it to zero (gray regions). {\bf Left:} combined fit in the $(2\text{--}\,6)$ {\rm keVee} energy window. {\bf Right:} fit of \LIBRA\ Phase 2 residuals in the $(1\text{--}\,3)$ {\rm keVee} energy window.}
\end{figure}

\begin{table*}[t]
\def\arraystretch{1.1}
\centering
\begin{tabular}{|c||cc||cc|}
\hline
Fitted & \multicolumn{2}{c||}{Fit to a  cosine modulation} & \multicolumn{2}{c|}{Fit to a secular
variation}\\ 
data & $\A$ [cpd/kg/keVee] & $\chi^2_{\rm cos}/{\rm d.o.f.}$ & $\B$ [cpd/kg/keVee/yr] & $\chi^2_{\rm saw}/{\rm d.o.f.}$ \\
\hlinewd{2pt}
\rowcolor{lightgray} DAMA/NaI & & & &  \\
\footnotesize{(2-4) keVee} & $0.0214\pm 0.0046$ & $36.3/36$ & $0.0783\pm 0.0141$  & $26.9/36$ \\
\footnotesize{(2-5) keVee} & $0.0200\pm 0.0037$ & $24.1/36$ & $0.0605\pm 0.0113$  & $24.7/36$ \\
\footnotesize{(2-6) keVee} & $0.0178\pm 0.0031$ & $36.9/36$ & $0.0602\pm 0.0094$   & $29.6/36$ \\
\hline
\rowcolor{lightgray}  LIBRA Phase I & & & &  \\
\footnotesize{(2-4) keVee} & $0.0164\pm 0.0022$ & $53.5/49$ & $0.0452\pm 0.0059$   & $51.9/49$ \\
\footnotesize{(2-5) keVee} & $0.0120\pm 0.0016$ & $42.8/49$ & $0.0302\pm 0.0044$   & $50.5/49$ \\
\footnotesize{(2-6) keVee} & $0.0095\pm 0.0013$ & $30.0/49$ & $0.0249\pm 0.0035$   & $33.6/49$ \\
\hline
\rowcolor{lightgray}  LIBRA Phase II & & & &  \\
\footnotesize{(1-3) keVee} & $0.0182\pm 0.0023$ & $61.2/51$ & $0.0475\pm 0.0062$   & $67.1/51$ \\
\footnotesize{(1-6) keVee} & $0.0103\pm 0.0011$ & $52.0/51$ & $0.0230\pm 0.0029$   & $83.7/51$ \\
\footnotesize{(2-6) keVee}  & $0.0093\pm 0.0011$ & $44.8/51$ & $0.0197\pm 0.0030$   & $72.9/51$ \\
\hlinewd{2pt}
\rowcolor{lightgray}  LIBRA  I and II   & & & &  \\
\footnotesize{(2-6) keVee} &  $0.0094\pm 0.0008$ & $74.8/101$ & $0.0219\pm 0.0026$  & $107.7/101$  \\
\hline
\rowcolor{lightgray}  DAMA combined & &   &  & \\
\footnotesize{(2-6) keVee} & $0.0100\pm 0.0008$ & $118.8/138$  & 
$0.0240\pm 0.0022$
& $152.9/138$ \\
\hline
\end{tabular}
\caption{\label{tab:fittoDAMA} \em Best-fit values of the DM cosine amplitude $\A$ and the sawtooth coefficient $\B$ obtained by fitting the \DAMA\  residuals  in different energy intervals, together with the corresponding $\chi^2/\text{d.o.f}$. Our fits of the cosine amplitudes agree with those reported by \LIBRA.}
\end{table*}

Keeping in mind that both the cosine and sawtooth provide acceptable fits to the data, we compare the two hypotheses by means of a likelihood ratio test.
Indeed, when fitting data with many degrees of freedom,
a likelihood-ratio test is a more powerful statistical indicator than the $\chi^2$ test.
The two tests answer different questions: the $\chi^2$ compares
the considered model to a generic model, such that
all statistical fluctuations contribute.
On the other hand, the $\Delta\chi^2$ compares two specific models 
such that only those statistical fluctuations that discriminate among the 
two models contribute.
We perform a $\Delta\chi^2$ fit of the data to a generic superposition of a sawtooth plus a cosine. 
We fix the period of the cosine to one year, and its peak to June 2nd, as predicted for a DM-induced annual modulation signal, so that the only free parameter is the amplitude $\A$.
The sawtooth also has one free parameter, the slope $\B$, since its phase and duration are fixed by the \DAMA\ analysis choices.
Fig.\fig{DAMAfit_AvsB} shows the results of the fit as 68\%, 95\% and 99.7\% allowed contours of $\A$ and $\B$.  On the left we consider the $(2\text{--}\,6)$ keVee energy interval for the three phases of \DAMA, while on the right we fit data in the lower energy window ($(2\,\text{--}\,4)$ keVee for both \NaI\ and \LIBRA\ Phase~1, and $(1\text{--}\,3)$ keVee for \LIBRA\ Phase~2). 
In the $(2\,\text{--}\,6)$ keVee interval the earlier \NaI\ data resemble more a sawtooth, while the more precise 
later Phase~2 data favour a cosine-dominated fit. 
Data in the lower energy interval are fitted equally well by both possibilities, a cosine or a secular variation.

Finally, we consider the cosine phase as an additional free parameter, still keeping the period fixed to one year, and quantify how much a possible sawtooth component relaxes the determination
of the cosine amplitude and phase.
In the left panel of fig.\fig{DAMAfit_AvsPhi} we show the regions of $\A$ and $t_{\rm peak}$
preferred at
68\%, 95\% and 99.7\% C.L.
by the combined \DAMA\ data-set in the $(2\text{--}\,6)$ keVee energy interval, both allowing for a free sawtooth component or setting it to zero. In the right panel we show the same but only for \LIBRA\ Phase~2 in the $(1\text{--}\,3)$ keVee energy interval. 
The best-fit regions assuming zero sawtooth (in gray)
reproduce  the results reported by \DAMA~(see {\it e.g.}\ fig.~15 of~\cite{1805.10486}). 
Allowing for a sawtooth component substantially affects the best fit regions.
In particular, the phase of the cosine becomes more uncertain and can more significantly differ from
the value characteristic of a DM signal, especially 
when fitting data at lower energy.

\bigskip 

The \DAMA\ collaboration analyzed the data by adopting two additional methods that are potentially relevant to indirectly clarify the time-dependence of the background.
In~\cite{0804.2741,1308.5109,1805.10486} a frequency analysis of the residuals has been presented. In section~5 of~\cite{1805.10486} this result is complemented at frequencies below about 1/yr with a spectral analysis of the yearly averaged total rate. We have summarized relevant details of the procedure at the end of our section~\ref{MC}.
We report in the right panel of fig.~\ref{fig:power} the power spectrum of the combined \LIBRA\ Phase~1 and Phase~2 data, taken from~\cite{1805.10486}, as black and red curves for the high and low frequencies, respectively.
This power spectrum can be compared to what expected from a sawtooth or cosine annual modulation (shown in the left panel of fig.~\ref{fig:power}). The green region in the figure shows the global 90\% C.L. upper limit obtained from Monte Carlo simulations, assuming a sawtooth with coefficient $\B$ as in our best fit to \LIBRA\ data of eq.~\eqref{sys:Bfit}. One can see that both the sawtooth and cosine hypotheses are compatible with the experimental data, reproducing the peak
at a frequency of about 1/yr, without giving significant extra power at different frequencies.

In addition, a direct fit of signal plus background to the total event rate has been performed (see {\it e.g.}\ section~6 of~\cite{1805.10486}). In this case, however, the constant part of the background --- $b_{jk}$ in the notation of~\cite{1805.10486} --- is left free within each cycle~\cite{private}. This is effectively equivalent to fitting residuals, and does not avoid the issues discussed here.

\begin{figure}[t]
\includegraphics[height=4.1cm]{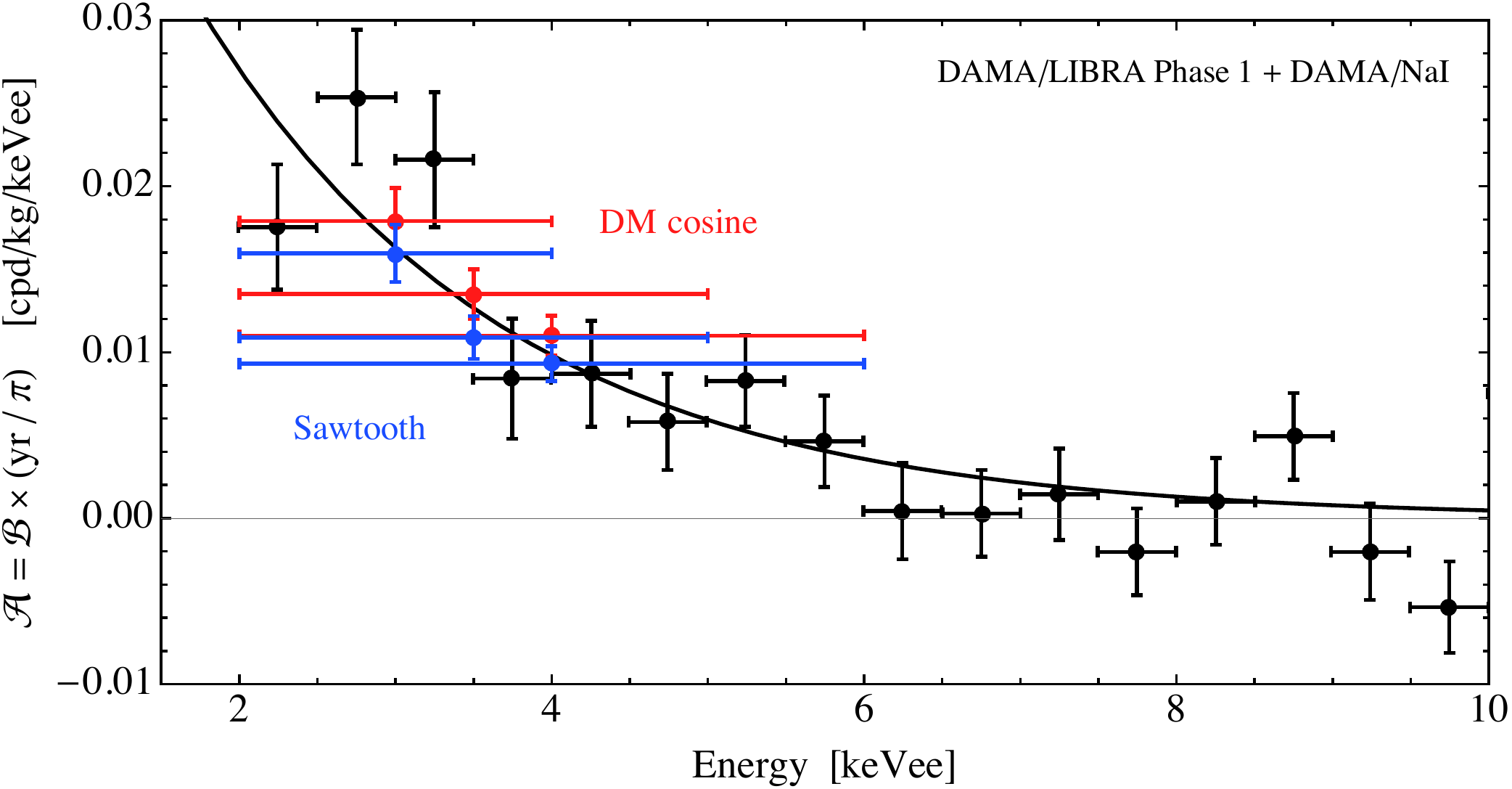}\hfill%
\includegraphics[height=4.1cm]{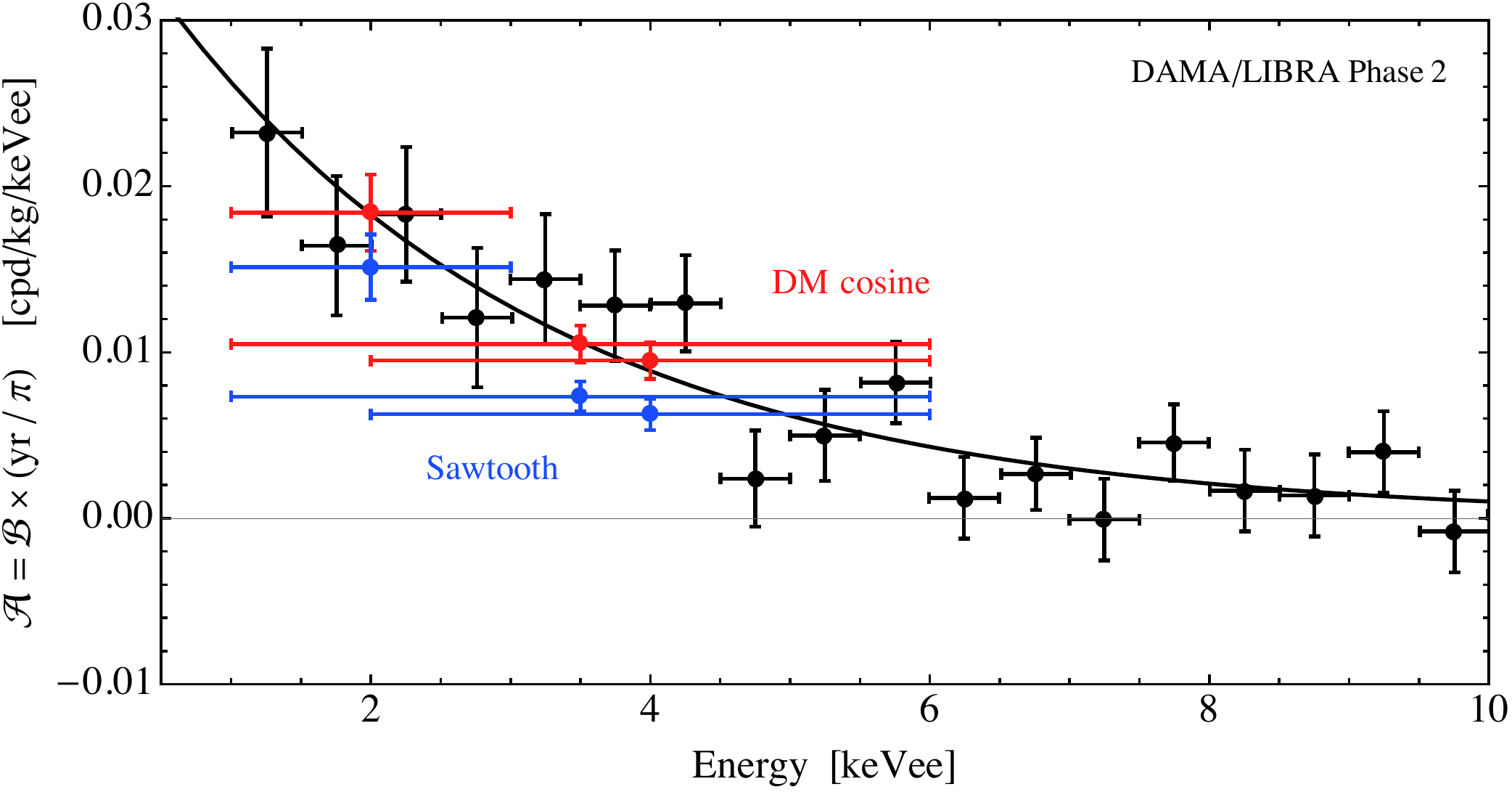}
\caption{\label{fig:DAMABvsAenergy}
\em Energy dependence of the modulated signal observed by \DAMA. The black points are the amplitudes of the cosine signal reported by DAMA in {\rm 0.5 keVee} energy bins~\cite{1805.10486}; for comparison we also show the DM cosine amplitudes in the wider energy intervals discussed above (red points), and our best-fit values for a linearly growing background (blue points). The black curves fit the data assuming an exponential energy dependence. {\bf Left:} Combination of \LIBRA\ Phase~1 and \NaI. {\bf Right:} \LIBRA\ Phase 2.}
\end{figure}

\subsection{Energy dependence of the secular variation}\label{fitE}

We now discuss the possible energy spectrum of the background rate
$R_{0}(t)$ of eq.~\eqref{eq:Rlinear}. 

The DAMA collaboration provided the energy dependence of the cosine signal amplitude (see for example fig.~10 of~\cite{1805.10486}), showing separately the results for \LIBRA\ Phase~2, and for the combined \NaI\ and \LIBRA\ Phase~1. We report these data in the two panels of fig.~\ref{fig:DAMABvsAenergy} in black. The decrease of the rate with energy is approximately exponential as expected for DM-induced recoils.
Above 6 keVee the modulated signal amplitude is compatible with zero.

We cannot repeat our fit over the whole energy range to extract the full energy dependence of the $\B$ coefficient, since the detailed time-dependence of the modulation is available only in the few energy bins discussed above. We however notice that in these three energy bins the values of $\B$ determined by fitting the data with a sawtooth, if compared with the cosine fit of \DAMA, are close to the na\"ive expectation $\B \simeq \A \pi/T$ (see eq.\eq{ABequiv}). This is shown in fig.~\ref{fig:DAMABvsAenergy}, where the red points represent the amplitude of the modulated rate reported by \DAMA, while the blue points correspond to our best-fit values of the sawtooth coefficient (rescaled by $\pi/\text{year})$. As one can see, the two sets of points are roughly compatible with each other, especially for the earlier experimental phases, and in the lower energy bins. We therefore use the detailed energy spectrum provided by DAMA (black points in the figure) as an approximate estimate of the energy dependence of $\B$.\footnote{Since data is provided only for the combined \NaI\ and \LIBRA\ Phase~1 datasets, we consider a single $\B$ for the two setups, as opposed to what was done in table~\ref{tab:fittoDAMA}.}

\begin{figure}[t]
\includegraphics[height=4.9cm]{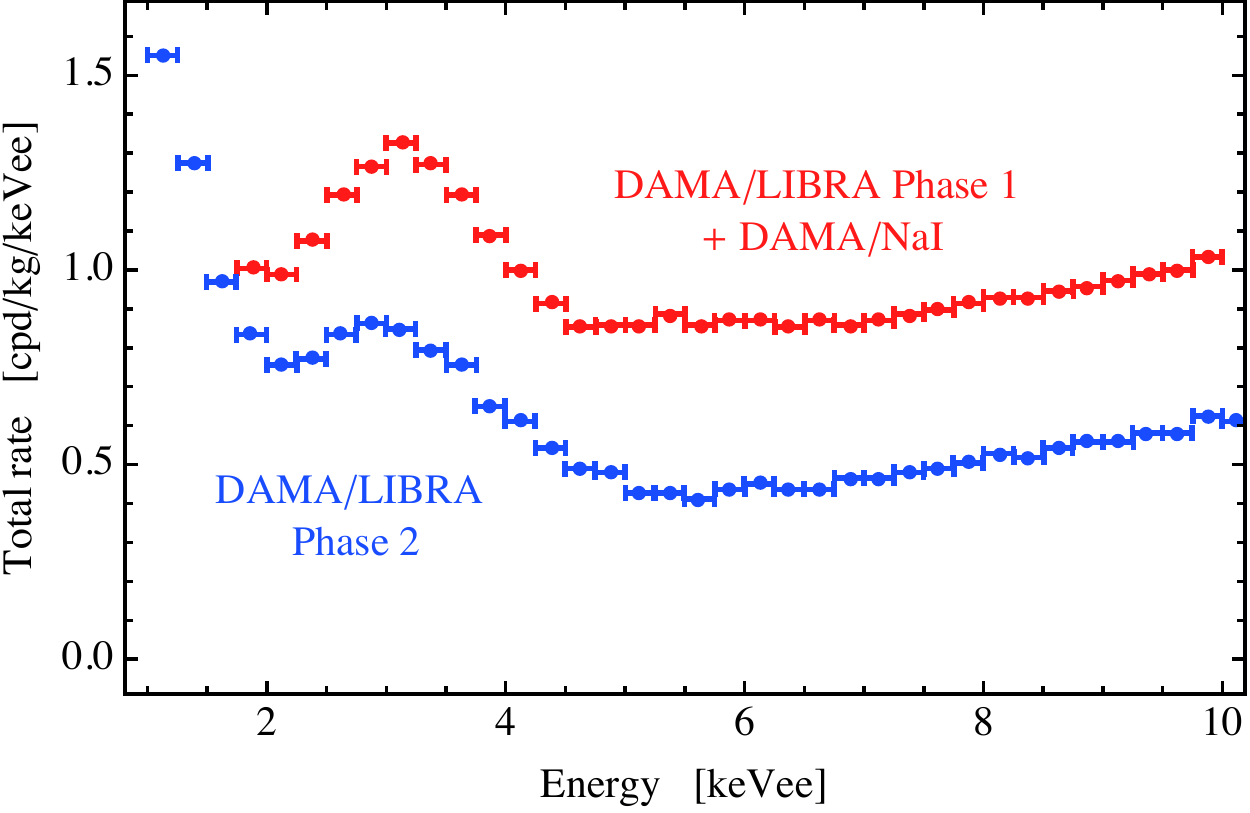}\hfill%
\includegraphics[height=4.9cm]{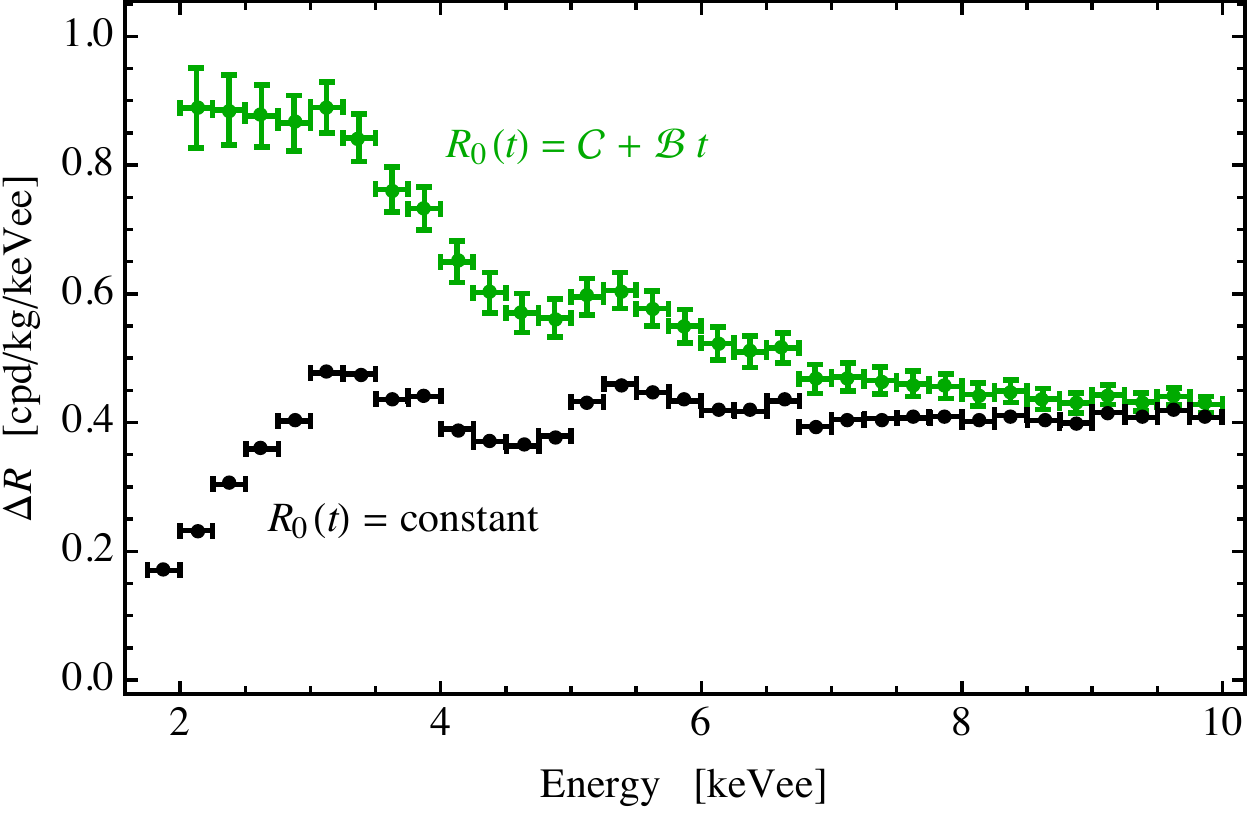}
\caption{\label{fig:totalEnergy}
\em {\bf Left:} Rate  averaged over all data-taking time
for \LIBRA\ Phase 2 (blue points), and for \LIBRA\ Phase 1 and {\rm \NaI} combined (red points). {\bf Right:} Difference in the total rate between the end of Phase 1 and the beginning of Phase 2, calculated assuming both a constant rate (black points), and a background that linearly grows with time (green points). In the latter case the values of $\B$ and $\C$ are chosen in order to reproduce the modulated signal and the cumulative rates; the errors come from the uncertainty in the residuals.}
\end{figure}
 
The constant part of the background $\C$ is determined by the energy spectrum of the total rate averaged over the whole data-taking period,
shown in the left panel of fig.~\ref{fig:totalEnergy}. 
These cumulative total rates differ between \LIBRA\ Phase 1 (red points taken from fig.~1 of~\cite{0804.2741}) and Phase 2 (blue points taken from fig.~1 of~\cite{1805.10486}), 
with a reduction of roughly a factor two between the two phases, most likely due to some improvement in the experimental setup. 
In the absence of a secular term in the background, 
this implies a change in the constant $\C$ which is trivially given by the difference of the cumulative rates.
If a slowly growing term $\B t$ is present, by averaging over the data-taking period
one gets that the variation of the total rate between the end of Phase~1 and the beginning of Phase~2 is given by
\beq
\Delta R\equiv R_{0}(t_{\rm 1,end}) - R_{0}(t_{\rm 2,start}) = 
R_{\rm 1} - R_{\rm 2} + \frac{\B}{2}(\Delta T_{\rm 1} + \Delta T_{\rm 2})\,,
\eeq
where $R_{\rm 1,2}$, and $\Delta T_{\rm 1,2}$ are the average rate and total duration of the two phases. Since 
the energy spectra of $R_{\rm 1,2}$ and $\B$ are different, we report $\Delta R$ as a function of recoil energy in fig.~\ref{fig:totalEnergy} (right) 
for the two cases where a secular term is absent (black), or is fitted from the modulated signal (green). We do not give a physical explanation on the energy dependence of $\Delta R$. 

Two comments are in order. First, the change in the averaged rate between the phases could in principle be entirely due to the secular term, if $\B<0$. A slowly decreasing background, however, would produce an oscillation peaked on December, in counterphase with the observed modulation. Second, the high-energy tails of the black points in fig.~\ref{fig:totalEnergy} (right) are approximately constant; one could therefore wonder whether the full difference could be explained by the energy dependence of the secular term $\B$ plus a constant shift $\Delta R$. This is indeed possible for a positive $\B$ with a spectrum peaked at low energies (below $\sim 5$ keVee), but its value in this case would be too small to completely reproduce the modulated signal with a sawtooth.

\subsection{Possible slowly-varying backgrounds}\label{bckt}
A slow variation of the rate could, in principle, be produced by some signal or by some background.
We here elaborate on the second possibility.
The background in underground detectors in the keV energy range has spectral features and time behaviours that are poorly understood. We first consider possible slowly growing backgrounds, as their presence would give 
an apparent \DAMA\ modulation  peaked around June, as shown above.
Various possible effects may create an increasing background, 
especially in a fixed energy window around few keV, 
as in the Region of Interest (RoI) of the \DAMA\ experiment.  We list  a few generic examples of possible increasing backgrounds, not necessarily related to the specific case of study:
\begin{itemize}
\item[$\diamond$] Increase due to out-of-equilibrium physical effects.  A classical example, although relevant at different energies, is the increase of the $^{210}$Po,  from a $^{210}$Pb initial contamination in the transient phase after a broken equilibrium in the contamination chain Pb $\to$ Bi $\to$ Po. An increase of the backgrounds in NaI crystals could also arise from the diffusion of dirty materials from the surface into the bulk, as well as  external gas cumulative contamination, as {\it e.g.} $^{222}$Rn. Other relevant  out-of-equilibrium physical effects related to an increase of the backgrounds might be possible and  can in principle be tested by dedicated analyses. 
A quantitative study of the backgrounds in NaI crystals has been reported in several papers by \DAMA\ (see~\cite{1306.1411} and references therein) and more recently by other collaborations ({\it e.g.}~\cite{1703.01982}). 
\item[$\diamond$] Increase due to instrumental effects. These are related to the photo-multipliers (PMT) and electronics, and they are more difficult to isolate with respect to the previous ones. For example, helium gas permeating through PMT glass increases the after-pulse rate~\cite{bib:pmt} creating fake low energy signals simulating the real scintillation. These are not easily rejectable through simple pulse shape discriminations, as  suggested in \cite{1901.02139} (although with completely different implications that we do not consider in this work). Another example is the possible increase of the electronic noise leaking in the acceptance region after pulse shape discrimination due to a generic loss of performance of the hardware.
\item[$\diamond$]  Apparent increase due to a general degradation of the detector resolution.
This is potentially relevant in a fixed energy window close to the software threshold. Here the limit conditions of the reconstruction performances in terms of efficiency and reliability are usually reached.  
\end{itemize}

We next consider the opposite scenario where the background is  slowly decreasing with time. In this case the analysis procedure adopted by \DAMA\ would transform the background into a sawtooth that mimics an annual modulation peaked around December, as discussed in section~\ref{modsec}. 
Then, a decreasing background cannot explain the annual modulation reported by \DAMA. However, the sawtooth
could destructively interfere with additional periodic signals, in particular possibly shifting their phase,
and a real DM signal would be distorted and attenuated.

A decrease of the total background rate could arise from different radioactive contaminants present in NaI crystals at the beginning of the data-taking as short and medium
term cosmogenic activations or intrinsic out-of-equilibrium contaminants \cite{1812.01377,1812.01472,1905.12861,1903.03973,1910.03782}. The most critical  ones are those with a half-life $T_{1/2}$  of a few tens of years, comparable to the total time of data-taking, and an energy spectrum in the RoI of the detector. Examples of possible decreasing backgrounds in NaI crystals in the keV energy range~include:
\begin{itemize}
\item Decay of relatively short-lived isotopes. For example, $^{109}$Cd with $T_{1/2} = 463 $ days and $^{22}$Na with $T_{1/2} = 2.6 $ yr. These contaminants are no longer relevant for \DAMA\ because they have fully decayed after few years of data acquisition. 
\item Decay of medium-lived isotopes  with a time-scale of order $\sim$10 yr. Typical contaminants that could be present at the beginning of  data-taking are the cosmogenic $^{3}$H with $T_{1/2} = 12.3 $ yr, and $^{210}$Pb with $T_{1/2} = 22.3 $ yr, which have about 40\% and 16\% of the energy spectrum falling in the RoI, respectively. In particular, the tritium has a $\beta$-spectrum peaked around 3 keV and with an endpoint of 18.6 keV. Being produced by cosmogenic activation, this contaminant is present in NaI crystals (grown in above-ground facilities) proportionally to the time exposure between their production and installation underground~\cite{1706.05818}.  
\end{itemize}


Let us stress that here we do not want to make any claim about the presence or absence of a given background, and even less to provide an explanation for the \DAMA\ total rate. Since a careful reconstruction of the rate is strongly hampered by the poor knowledge of the background details, it would be extremely  useful to directly check whether the \DAMA\ background underwent a secular evolution.
The \COSINE\ (see {\it e.g.}~\cite{COSINE}) and \ANAIS\ (see {\it e.g.}~\cite{1903.03973}) experiments recently started data-taking using NaI crystals, and reported the total rates of each crystal as a function of time. 
The rates of these new experiments show, in their starting phase, a significant time-dependence.


\section{Conclusions}\label{concl}

The \DAMA\ collaboration reported an annual modulation in the single-hit scintillation events 
that peaks around June, 2$^{\rm nd}$, as predicted by a DM signal.
The statistical significance of the \DAMA\ signal is large, roughly $13\,\sigma$,
such that the annual modulation could be in principle visible 
even by simply plotting the event rate binned as a function of time.

However, the \DAMA\ collaboration reported residuals computed by subtracting the weighted average of 
its rate over cycles of roughly one year. 
This procedure can be dangerous, because a possible secular variation of the total rate would be transformed into a sawtooth with period coinciding with the duration of the \DAMA\ analysis cycles.
As a result, a slowly time-dependent rate, even if not oscillating, becomes a possible source of apparent modulation.
Within the statistical accuracy, the modulation possibly produced by the procedure
can resemble a cosine with period of one year.
Since the data-analysis cycles usually start around September, 
the apparent cosine would peak at the beginning of June if the background slowly grows with time.

In section~\ref{DAMA} we considered the most extreme possibility: a slowly and steadily growing background that fully reproduces the observed modulation amplitude in \DAMA. By means of a Monte Carlo simulation, we have shown that the DM signal can be mimicked, within statistical errors, by
a growth of the total rate on decennial time-scale.
A rate that decreases with time, on the other hand, would suppress the oscillating signal.

The \DAMA\ collaboration so far has not presented data about the time-dependence of its total rate,
that can confirm or exclude the presence of a slowly changing background.
We therefore tried to clarify the issue through the available indirect means in section~\ref{fit}.
The detailed time dependence of the published residuals can be fitted by an irregular sawtooth with a $\chi^2/{\rm d.o.f.}\approx 1$,
see fig.\fig{DAMAdata_26} and table~\ref{tab:fittoDAMA}. The available data are consistent with the modulation being interpreted as an artefact due to a yearly growth of the total rate of a few percent.
A DM cosine provides a somehow better fit to parts of the data-set --- specifically the higher energy bins of DAMA/LIBRA Phase 2 --- therefore preferring a DM-like signal as a dominant explanation of the observed modulation. A time-dependent background rate would have an impact on the signal also in this case, since it would affect the extraction of the cosine amplitude and phase from the data, as shown in fig.\fig{DAMAfit_AvsPhi}.
This effect can be important since the position of the sinusoid peak 
plays an important role in the interpretation of the modulation as a DM signal.
A spectral analysis of the residuals and of the annual averaged rate is found to be consistent 
with a slow variation of the rate, as well as with a true annual modulation.

Finally, in section~\ref{fitE} we discussed the energy spectrum 
of the possible slowly-varying component of the rate, which must be peaked at low energy, and in section~\ref{bckt} we speculate about possible origins of such a background. We do not attempt at giving any realistic explanation of a possible time-dependence of the \DAMA\ total rate; we rather just mention a few simple physical and instrumental effects that could produce a background that grows with time in the \DAMA\ region of interest.
It is worth mentioning that the \ANAIS\ and \COSINE\ experiments recently
started data-taking with NaI crystals similar to the \DAMA\ ones, and 
presented their rates as function of time, finding (in their initial phase) a sizeable time dependence~\cite{1903.03973,COSINE}.

In conclusion, from the data available to us,
we could not exclude the possibility that
the cosine modulation claimed by DAMA
is biased by a slow variation in the rate,
possibly due to some background or signal.
The most estreme possibility of no cosine modulation seems disfavoured,
but we could not safely exclude it.
More in general, a slowly increasing or decreasing rate would partially enhance or mask a true annual modulation, respectively.

The potential significance of our observation could, of course, be directly clarified by the measurement
of the \DAMA\ total rate as function of time. The absence of an appreciable variation of the rate over the time-scale of the experiment would exclude the relevance of the effects discussed in this paper for the \DAMA\ analysis.
A second possibility to test our observations would be to recompute the residual rate following a different procedure, such as subtracting the average background choosing cycles with different starting points or different durations. 
Even better, the above issues would be avoided  
by performing a detrend with continuous modelling of the possible time dependence of the background --- such as a polynomial fit or interpolation --- rather than a discontinuous average.
Another possibility is to perform a signal processing in which, along with the original time series, both power spectrum and phase information are shown.

The direct detection of Dark Matter interactions in underground laboratories is one of the main challenges of 
contemporary particle physics. The observation of a DM signal would be an important discovery, so that any alternative interpretation needs to be explored with  care. We therefore hope that an effort will be put by the experimental collaborations to settle the potential issues presented here.


\footnotesize

\subsubsection*{Acknowledgements}
This work was supported by ERC grant NEO-NAT and MIUR contracts 2017FMJFMW~and 2017L5W2PT (PRIN).\ 
We thank Sabine Hossenfelder,
Aldo Ianni, Marcello Messina and Tobias Mistele~for~discussions.



\end{document}